\documentclass[11pt,aps,prd,preprint,preprintnumbers,showpacs,groupedaddress,floatfix]{revtex4}

\usepackage{graphicx}
\usepackage{dcolumn}
\usepackage{bm}

\begin{document}

\begin{flushright}{\bf{UK/03-03}}\end{flushright}

\title{Chiral Logs in Quenched QCD}

\author{Y. Chen$^{a,b}$, S.J. Dong$^{a}$, T. Draper$^{a}$, I. Horv\'{a}th$^{a}$, \mbox{F.X. Lee$^{c,d}$,} 
K.F. Liu$^{a}$, N. Mathur$^{a}$ and J.B. Zhang$^{e}$}

\affiliation{%
\centerline{$^{a}$Department of Physics and Astronomy, University of Kentucky, Lexington, KY 40506}
\centerline{$^{b}$Institute of High Energy Physics, PO Box 918, Beijing 100039, China}
\centerline{$^{c}$Center for Nuclear Studies, Department of Physics,} 
\centerline{George Washington University, Washington, DC 20052}
\centerline{$^{d}$Jefferson Lab, 12000 Jefferson Avenue, Newport News, VA 23606}
\centerline{$^{e}$CSSM and Department of Physics and Mathematical Physics,}
\centerline{University of Adelaide, Adelaide, SA 5005, Australia}%
}

\begin{abstract}
The quenched chiral logs are examined on a $16^3 \times 28$ lattice with
Iwasaki gauge action and overlap fermions.  The pion decay constant $f_{\pi}$
is used to set the lattice spacing, $a = 0.200(3)\,{\rm fm}$.  With pion mass
as low as $\sim 180\,{\rm MeV}$, we see the quenched chiral logs clearly in
$m_{\pi}^2/m$ and $f_P$, the pseudoscalar decay constant.  We analyze the data
to determine how low the pion mass needs to be in order for the quenched
one-loop chiral perturbation theory ($\chi$PT) to apply.  With the constrained
curve-fitting method, we are able to extract the quenched chiral log parameter
$\delta$ together with other low-energy parameters.  Only for $m_{\pi} \leq
300\,{\rm MeV}$ do we obtain a consistent and stable fit with a constant
$\delta$ which we determine to be 0.24(3)(4) (at the chiral scale
$\Lambda_{\chi}=0.8\,{\rm GeV}$).  By comparing to the $12^3 \times 28$
lattice, we estimate the finite volume effect to be about 2.7\% for the
smallest pion mass. We also fitted the pion mass to the form for the re-summed
cactus diagrams and found that its applicable region is extended farther than the range
for the one-loop formula, perhaps up to $m_{\pi} \sim 500-600$ MeV. The scale 
independent $\delta$ is determined to be 0.20(3) in this case.  We study the 
quenched non-analytic terms in the nucleon mass and find
that the coefficient $C_{1/2}$ in the nucleon mass is consistent with the
prediction of one-loop $\chi$PT\@.  We also obtain the low energy constant
$L_5$ from $f_{\pi}$. We conclude from this study that it is imperative to
cover only the range of data with the pion mass less than $\sim 300\,{\rm MeV}$
in order to examine the chiral behavior of the hadron masses and decay
constants in quenched QCD and match them with quenched one-loop $\chi$PT\@.
\end{abstract} 
\pacs{11.15.Ha, 12.28.Gc, 11.30.Rd}
\maketitle
\thispagestyle{empty}

\section{Introduction}

One of the important goals of lattice QCD is to understand, from first
principles, low-energy hadron phenomenology as a consequence of chiral
symmetry.  However, there have been problems with regard to chiral symmetry
since the advent of the lattice formulation.  The lack of chiral symmetry in
traditional approaches (such as Wilson fermions and staggered fermions) leads
to a plethora of conceptual and technical difficulties.  In particular, due to
critical slowing down and the existence of exceptional configurations, there is
a practical limit on how low in quark (or pion) mass one can carry out a Monte
Carlo simulation.  The existing calculations with these fermions are typically
limited to pion mass greater than $\sim 300\,{\rm MeV}$.  As such, the
extrapolation from the current lattice data with pion mass above $\sim
300\,{\rm MeV}$, sometimes well above, to the physical pion mass of $137\,{\rm
MeV}$ is both a theoretical and a practical challenge.  The primary concern is
whether the quark mass region (of the lattice data) is inside the range where
$\chi$PT applies.  If not, one does not have a reliable analytic form in quark
mass dependence to allow for an unbiased extrapolation.  Furthermore, most of
the quenched $\chi$PT calculations are carried out at one-loop order; this
limits its range of applicability further, to even smaller quark masses.  How
small does the quark mass have to be in order to see the chiral behavior
predicted by $\chi$PT\@?  Is the strange quark in the radius of convergence of
$\chi$PT\@?  Is there a way to model the chiral extrapolation from well above a
$300\,{\rm MeV}$ pion mass without introducing systematic errors?  These were
some of the issues addressed in the panel discussion during the {\sl Lattice
2002\/} conference~\cite{ber02}.

Fortunately, recent progress in lattice fermions has led to formulations (such
as domain-wall fermions, overlap fermions, and the fixed-point actions) with
lattice chiral symmetry at finite cut-off.  As a consequence, many
chiral-symmetry relations~\cite{neu01,nn93,nn95,neu98,lus98} and the quark
propagator~\cite{liu02} preserve the same structure as in the continuum.  While
a lot of progress has been made in checking the chiral symmetries in all these
three formulations, we find that overlap fermions~\cite{neu98} offer several
distinct advantages.  For example, it has been demonstrated that one can
simulate near the physical quark-mass region with very gentle critical slowing
down~\cite{dll00}.  Secondly, the overlap-fermion inverter can accommodate
multiple masses in the calculation of quark propagators~\cite{ehn99b} and thus
is particularly well-suited for studying the details of the
logarithmically-varying mass dependence of physical observables.  Thirdly, we
find that the $O(a^2)$ and $O(m^2 a^2)$ errors are
small~\cite{liu02,dll00,ddh02}. (There are no $O(a)$ errors due to chiral
symmetry~\cite{knn97}.)  For these reasons, in this paper we use overlap
fermions in a quenched lattice calculation to study the chiral logs in the pion
mass, the pseudoscalar decay constant $f_P$, the pion decay constant $f_{\pi}$,
and the nucleon mass.  We determine how small a quark mass or pion mass needs
to be for one-loop $\chi$PT to be valid.  In the present work, we only consider
hadrons with degenerate quark masses.  Quantities involving the strange quark
such as the kaon mass and decay constant will be studied later.

\section{Numerical details}

Our calculation is done on a $16^3 \times 28$ lattice with 80 quenched gauge
configurations generated with the Iwasaki gauge action~\cite{iwa85} at $\beta
=2.264 $ and the quark propagators are calculated with overlap fermions.  The
lattice spacing is $0.200(3)\,{\rm fm}$, as determined from the pion decay
constant $f_{\pi}(m_{\pi})$; the box size is $3.2\,{\rm fm}$.
  
The massive overlap Dirac operator~\cite{neu98a} is defined so that at
tree-level there is no mass or wavefunction renormalization~\cite{ddh02},
\begin{equation} \label{over-op}
D(m) = \rho + \frac{m}{2} + (\rho - \frac{m}{2} ) \gamma_5 \epsilon (H).
\end{equation}
Here $\epsilon (H) = H /\sqrt{H^2}$ is the matrix sign function and $H_W$ is
taken to be the hermitian Wilson-Dirac operator, i.e.\ $H = H_W = \gamma_5 D_{\rm
W}$.  Here $D_{\rm W}$ is the usual Wilson fermion operator, except with a
negative mass parameter $- \rho = 1/2\kappa -4$ in which $\kappa_c < \kappa <
0.25$.  We take $\kappa = 0.19$ in our calculation which corresponds to $\rho =
1.368$.

Throughout the paper, we shall use lattice units for dimensionful quantities,
except that the lattice spacing $a$ will be explicit in figures.

\subsection{Zolotarev Approximation}

The numerical aspects of the calculation have been given
elsewhere~\cite{dll00,ddh02}.  The new ingredient is the adoption of the
Zolotarev optimal rational approximation~\cite{vfl02,chh02} of the matrix sign
function.

To approximate the function $\frac{1}{\sqrt{x}}$, it is suggested~\cite{vfl02}
that one should consider the minimization of
\begin{equation} \label{Zolo}
|| 1 - \sqrt{x}f(x)||_{\infty}^{[\lambda_{min}^2, \lambda_{max}^2]},
\end{equation}
with $f(x)$ being approximated by a rational polynomial $\tilde{f}(x) \in
R^{m-1,m} = p(m-1)/q(m)$ where $p(m-1)$ and $q(m)$ are polynomials with degree
$m-1$ and $m$, respectively.  It was proven by Zolotarev~\cite{zol1877} that $x
\tilde{f}(x^2) \in R^{2m-1,2m}$ is the $||.||_{\infty}$-optimal rational
approximation for the sign function on $[-|\lambda_{max}|, -|\lambda_{min}|]
\cup [|\lambda_{min}|, |\lambda_{max}|]$.  By way of Zolotarev's
theorem~\cite{zol1877,idk00,pp87}, there is an analytic solution
\begin{equation}
\tilde{f}(x) = A \,\frac{\prod_{l =1}^{m-1}(x + c_{2l})}{\prod_{l =1}^{m}(x + c_{2l-1})},
\end{equation}
where the coefficients can be expressed in terms of Jacobian elliptic functions
\begin{equation}
c_l = \frac{sn^2(lK/2m;\kappa)}{1 - sn^2(lK/2m; \kappa)},\quad l = 1,2,\ldots,2m-1,
\end{equation}
where $\sqrt{1 - \kappa^2} = |\frac{\lambda_{max}}{\lambda_{min}}|$ and $K$ is
the complete elliptic integral.  According to Ref.~\cite{vfl02}, $A$ is
determined by the condition
\begin{equation}  \label{con_A}
\max_{C[1,(\frac{\lambda_{\mbox{\tiny max}}}{\lambda_{\mbox{\tiny
        min}}})^2]}[1-\sqrt{x}f(x)]
=-\!\!\!\!\!\!\!\!
\min_{C[1,(\frac{\lambda_{\mbox{\tiny max}}}{\lambda_{\mbox{\tiny
        min}}})^2]}[1-\sqrt{x}f(x)].
\end{equation}

\begin{figure}
\includegraphics[angle=90,width=0.8\hsize]{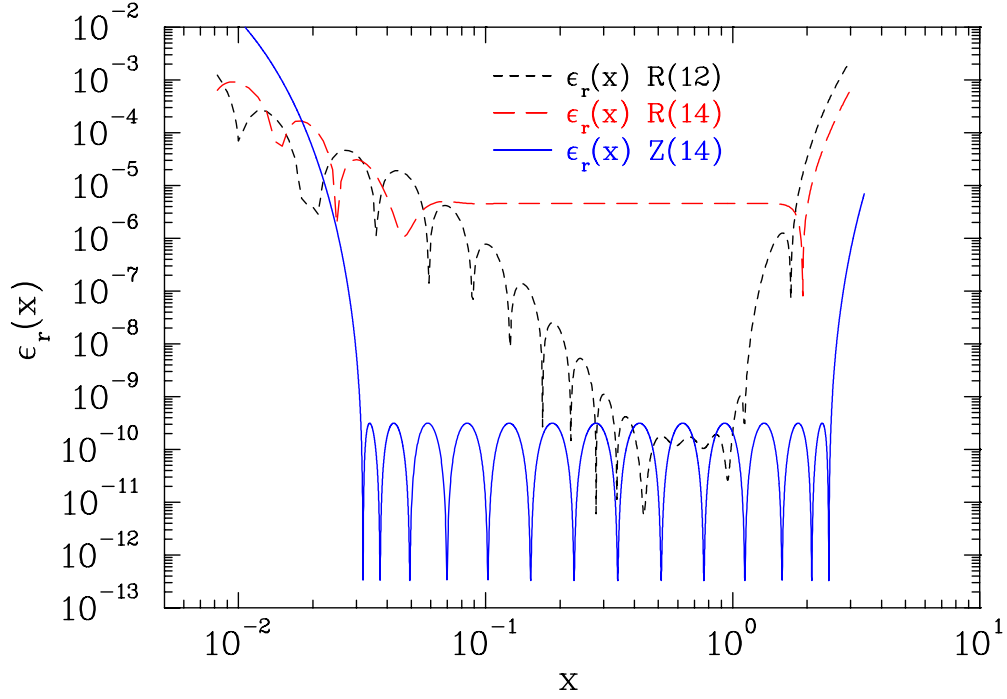}
\caption{\label{zolo-fig} The residual $\epsilon_r(x)$ of the approximation to
the sign function.  The solid line is the current 14th degree Zolotarev
approximation.  The dashed and dotted lines are the 14th~\cite{ehn99a} and
12th~\cite{dll00} degree Remez approximations respectively.}
\end{figure}

However, there is a worry when the sign function $\epsilon(H)$ is not bounded
by unity because of slight numerical imprecision.  From the Cauchy-Schwartz
inequality $||\gamma_5 \tilde{\epsilon}(H)|| < ||\gamma_5||\,
||\tilde{\epsilon}(H)||$, the overlap operator $D(0) = 1 + \gamma_5
\epsilon(H)$ with $\epsilon(H)$ approximated by $\tilde{\epsilon}(H)$ could
have superfluous unphysical zero modes~\cite{neu00b}.  In view of this, we
fine-tune the parameter $A$ by reducing it a little from the value determined
from Eq.~(\ref{con_A}) to minimize the positive residual.  This doubles the
worst residuals.

For the Wilson action kernel $H_W = \gamma_5 D_W$ inside the sign function, the
largest eigenvalue is around 2.43 and is fairly stable over the configuration
space.  Furthermore, we find that the eigenstates of $H_W$ to be projected out
become dense at $\sim 0.05-0.06$.  Thus, it is sufficient to specify $\sqrt{1 -
\kappa^2} = |\frac{\lambda_{max}}{\lambda_{min}}| = 80 $.  With
$|\lambda_{max}| = 2.5$, this gives $|\lambda_{min}| = 0.0315$ which is smaller
than the largest eigenstates ($\sim 0.05 - 0.06$) we project out from the
kernel $H_W$.  Since the coefficients of the rational polynomial approximation
are expressed in terms of the Jacobian elliptic functions, one can use as high
a degree of the polynomials as the memory can hold in the multi-mass algorithm.
In practice, we use 14th degree which is a sufficiently good approximation for
our calculation.  We plot the residual $\epsilon_r(x) = \frac{x}{\sqrt{x^2}} -
Z_{14}(x)$ for $x > 0$ in Fig.~\ref{zolo-fig} as a function of $x$.  Here
$Z_{14}(x)$ is the 14th degree Zolotarev approximation $x\tilde{f}(x^2)$ from
Eq.~(\ref{Zolo}).  As we can see, in the selected window [0.031, 2.5], the
approximation is better than $3.3 \times 10^{-10}$.  This is a good deal better
than the earlier attempts which approximate $g(x)$ to $\frac{1}{\sqrt{x}}$ with
the Remez algorithm in which case the resultant approximation to the sign
function with $x g(x^2)$ is not optimal in the range $[-|\lambda_{max}|,
-|\lambda_{min}|] \cup [|\lambda_{min}|, |\lambda_{max}|]$.  We plot the
residual $\epsilon_r(x)$ for the 14th degree~\cite{ehn99b} approximation
$R(14)$ and 12th degree~\cite{dll00} approximation $R(12)$ with the Remez
method in comparison with the 14th degree Zolotarev approximation $Z(14)$.  We
see that the Remez method gives residuals in the order of $10^{-5}$ in the
relevant range [0.031, 2.5].  We note that according to Chebyshev's theorem,
the errors of the optimal rational approximation $\tilde{f}(x) \in R^{m,n} =
p(m)/q(n)$ have $m + n + 2$ alternating signs in the range of the
approximation.  Since the Zolotarev approximation to the sign function is
optimal, it should satisfy this requirement.  In other words, there should be
29 alternating maxima and minima (including the endpoints of the range) for the
approximation with $m = 13, n = 14$.  As seen in Fig.~\ref{zolo-fig}, this
feature is preserved with the restriction allowing only positive residuals, as
described above.

\begin{table}[th]
\begin{center}
\caption{\label{zolo} Coefficients for the 14th degree Zolotarev approximation
in Eq.~(\ref{coeff}).}
\vspace*{0.2in}
\begin{tabular}{c@{\hspace{2em}}c@{\hspace{2em}}c} 
\hline
$i $& $c_i$&$q_i$ \\
\hline
  1  & 0.8371803911435546d$-$02  & 0.4203852824188996d$-$04 \\
  2  & 0.9813674433769438d$-$02  & 0.4230339571009471d$-$03 \\
  3  & 0.1294644813006436d$-$01  & 0.1458829357287322d$-$02 \\
  4  & 0.1831241561050564d$-$01  & 0.3894118671815802d$-$02 \\
  5  & 0.2684553134920763d$-$01  & 0.9481580995276351d$-$02 \\
  6  & 0.4004971283185919d$-$01  & 0.2225210303473774d$-$01 \\
  7  & 0.6031838397347246d$-$01  & 0.5146873615634705d$-$01 \\
  8  & 0.9155798451341547d$-$01  & 0.1185868564259925d$+$00 \\
  9  & 0.1406107013842408d$+$00  & 0.2742893835909246d$+$00 \\
 10  & 0.2211980404641135d$+$00  & 0.6437234073136943d$+$00 \\
 11  & 0.3673892837229880d$+$00  & 0.1567367648339766d$+$01 \\
 12  & 0.6933239005020095d$+$00  & 0.4183844803034055d$+$01 \\
 13  & 0.1812368256185377d$+$01  & 0.1442795672202632d$+$02 \\
 14  & 0.1555827970041009d$+$02  & 0.1451886134043600d$+$03 \\
\hline
\end{tabular}
\end{center}
\vspace*{-0.2in}
\end{table}


After partial fraction expansion, the approximated sign function in rational
polynomial $x \tilde{f}(x^2) \in R^{2m-1,2m}$ has the form
\begin{equation}   \label{coeff}
\frac{x}{\sqrt{x^2}} \sim \ x \sum_{i=1}^{N = 14} \frac{c_i}{x^2 + q_i}.
\end{equation}
The coefficients $c_i$ and $q_i$ for our 14th degree Zolotarev approximation
are listed in Table~\ref{zolo}.

In order to improve the approximation of the matrix sign function as well as
the convergence in the conjugate gradient inversion in the inner do
loop~\cite{ehn99b,dll00,ddh02}, it is desirable to project out the lowest
several eigenmodes of $H_W$.  We use the Iwasaki improved gauge action which
requires that a smaller number of eigenmodes be projected, compared to the
Wilson and L\"{u}scher-Weisz gauge actions~\cite{eh02}.  We calculated 140
small eigenvalues $\lambda$ and projected out the lowest $110$ -- $140$ eigenmodes
$\Psi_{\lambda}$, based
on the criterion that the residual of the eigenvalue, defined as $\mid\mid
H_W \Psi_{\lambda}-\lambda \Psi_{\lambda}\mid\mid/\mid\mid\lambda \sqrt{{\rm
dim.} (H_W)}\mid\mid$, be less then $10^{-7}$.  The next smallest eigenvalue
is $\sim 0.05$ -- $0.06$, which is well within the window where the
approximation for the sign function is better than $3.3 \times 10^{-10}$.  The
residuals of both the inner and the outer do loops for the inversion of the
massive overlap Dirac operator in Eq. (\ref{over-op}) are at the level of
$10^{-7}$.

\section{Pion mass and zero modes}  \label{pion_zero}

\begin{figure}[t]
\includegraphics[width=0.8\hsize]{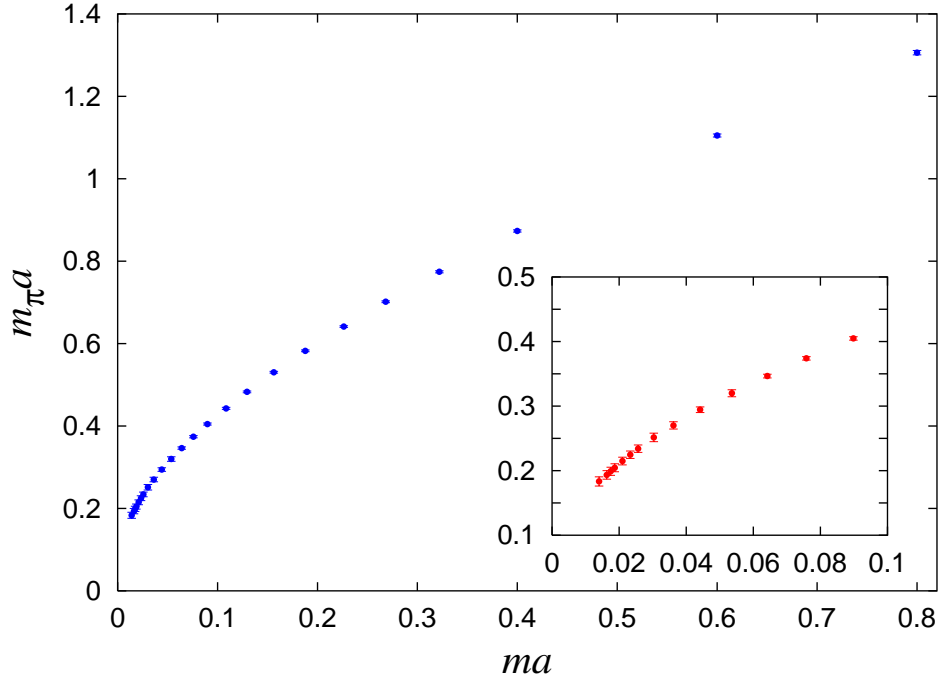}
\caption{\label{pi_08} The pion mass as calculated from the $\langle
P\,P\rangle$ correlator as a function of the bare quark mass $m$ from 0.014 to
0.8.  The insert is for the results with small quark masses up to $m=0.1$.}
\end{figure}

We look at the pion mass as calculated from the pseudoscalar correlator
$\langle\sum_{\vec{x}} P(\vec{x},t) P(0)\rangle$ where $P = \bar{\psi}i\gamma_5
(1 - D/2)\psi$.  The pion masses $m_{\pi}$ obtained from this correlator are
listed in Table~\ref{pion_mass} and plotted in Fig.~\ref{pi_08} as a function
of the quark mass $m$.  Our smallest pion mass is $182(8)\,{\rm MeV}$.

\begin{table}[ht]
\begin{center}
\caption{\label{pion_mass} 
Pion mass obtained from the $\langle P\,P\rangle$ correlator on a
$16^3\times 28$ lattice.  The pion decay constant $f_{\pi}$ is used to set the
lattice spacing, $a = 0.200(3)\,{\rm fm}$.}
\vspace*{0.2in}
\begin{tabular}{c@{\hspace{2em}}c@{\hspace{2em}}c@{\hspace{2em}}c} 
\hline
$m  $   & $m_{\pi}$  & $m_{\pi}\,{\rm (MeV)}$  & $\chi^2/N$  \\
\hline
0.80000 & 1.3106(42) & 1293(20) & 0.68 \\
		              
0.60000 & 1.1069(38) & 1092(17) & 0.60 \\
		              
0.40000 & 0.8740(30) & 862 (13) & 0.56 \\
		              
0.32200 & 0.7743(22) & 764 (12) & 0.54 \\
		              
0.26833 & 0.7017(24) & 692 (11) & 0.57 \\
		              
0.22633 & 0.6416(25) & 633 (10) & 0.63 \\
		           
0.18783 & 0.5829(28) & 575 (9)  & 0.72 \\
 		             
0.15633 & 0.5312(31) & 524 (8)  & 0.77 \\
		             
0.12950 & 0.4840(32) & 478 (8)  & 0.75 \\
 		             
0.10850 & 0.4438(36) & 438 (7)  & 0.70 \\
		             
0.08983 & 0.4050(38) & 400 (7)  & 0.64 \\
		             
0.07583 & 0.3748(38) & 370 (7)  & 0.62 \\
		             
0.06417 & 0.3470(40) & 342 (6)  & 0.58 \\
		             
0.05367 & 0.3200(43) & 316 (6)  & 0.53 \\
		             
0.04433 & 0.2940(42) & 290 (6)  & 0.55 \\
		             
0.03617 & 0.2693(48) & 266 (6)  & 0.52 \\
 		             
0.03033 & 0.2501(54) & 247 (6)  & 0.56 \\
		             
0.02567 & 0.2332(56) & 230 (7)  & 0.48 \\
		             
0.02333 & 0.2244(59) & 221 (7)  & 0.58 \\
		             
0.02100 & 0.2151(61) & 212 (7)  & 0.62 \\
		             
0.01867 & 0.2055(61) & 203 (7)  & 0.60 \\
		             
0.01750 & 0.2005(65) & 198 (7)  & 0.64 \\
		             
0.01633 & 0.1953(69) & 193 (7)  & 0.68 \\
		             
0.01400 & 0.1844(71) & 182 (8)  & 0.72 \\
\hline
\end{tabular}
\end{center}
\end{table}

\begin{figure}[t]
\includegraphics[width=0.8\hsize]{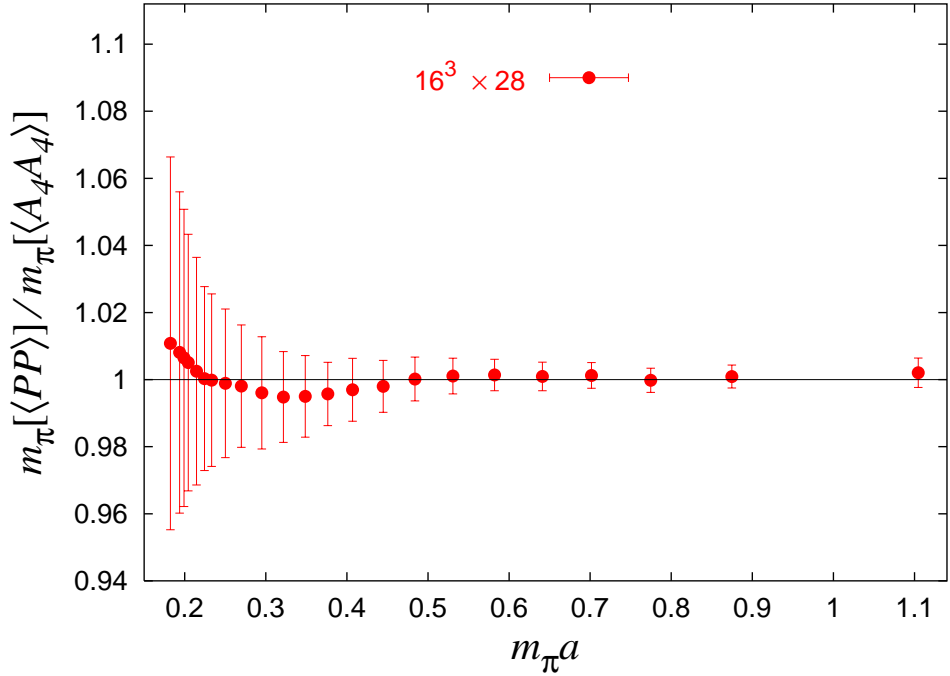}
\caption{\label{ratio_masses} Ratio of pion masses as calculated from the
$\langle P\,P\rangle$ and $\langle A_4\,A_4\rangle$ correlators as a function
of the $m_{\pi}$ obtained from the $\langle A_4\,A_4\rangle$ propagator.}
\end{figure}

\begin{figure}[h]
\includegraphics[width=0.8\hsize]{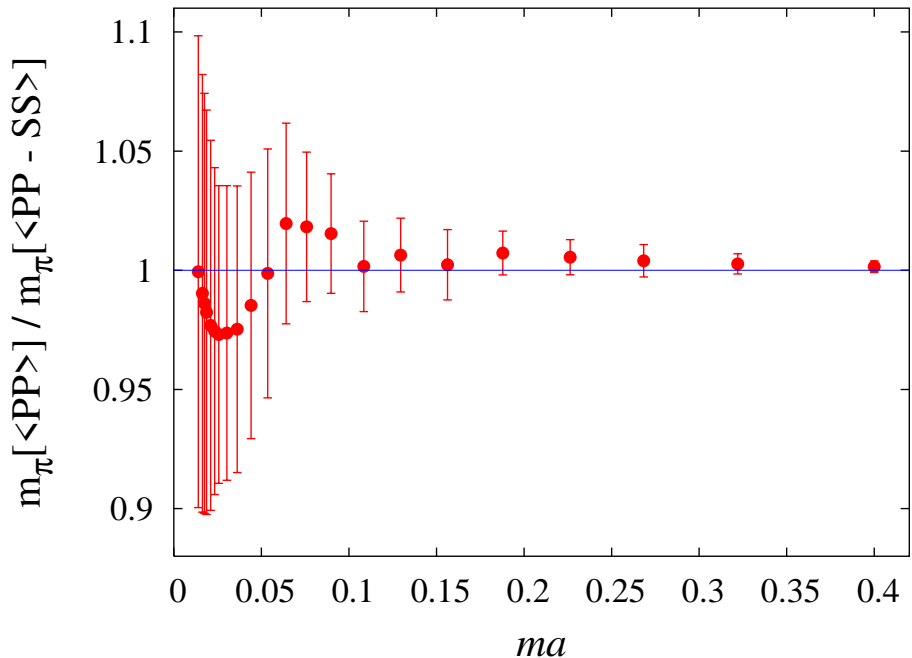}
\caption{\label{pp-ss} Ratio of pion masses as calculated from the
$\langle P\,P\rangle$ and $\langle P\,P\rangle - \langle S\,S\rangle$ correlators as a function
of the $m a$.}
\end{figure}

Consider the contributions of the zero modes to the meson correlator of the
interpolation field $M=\bar{\psi}\Gamma (1 - D/2)\psi$,
\begin{eqnarray}  \label{zero_meson}
\lefteqn{\int d^3x \,\langle M(x) M^{\dagger}(0)\rangle|_{zero\, modes} = }  \nonumber\\
& &-  \int d^3x \,{\biggl [}\sum_{i,j = zero\, modes}
\frac{\langle {\tt tr}(\psi^{\dagger}_j(x)\gamma_5 \Gamma\psi_i(x)){\tt tr}
(\psi^{\dagger}_i(0) \bar{\Gamma}\gamma_5\psi_j(0))\rangle}{m^2} \nonumber \\
& & + 2\sum_{i = 0, \lambda > 0}\frac{\langle {\tt tr}(\psi^{\dagger}_{\lambda}(x)\gamma_5
\Gamma \psi_i(x)){\tt tr}(\psi^{\dagger}_i(0)\bar{\Gamma}\gamma_5\psi_{\lambda}(0))\rangle}
{\lambda^2 + m^2}{\biggr ]},
\end{eqnarray}
where $\Gamma$ is the spinorial matrix and $\bar{\Gamma} = \pm
\Gamma^{\dagger}$.  For the pseudoscalar case with $\Gamma = \gamma_5$, the
zero modes contribute to both the direct and cross terms in the correlator.
They contribute also to the cross terms in the $\langle \sum_{\vec{x}}
A_4(\vec{x},t) P(0)\rangle$ and $\langle \sum_{\vec{x}} A_4(\vec{x},t) A_4(0)\rangle$ 
correlators, where $A_4 = \bar{\psi}i\gamma_4\gamma_5 (1 - D/2)\psi$, but not
to the direct terms. In view of the worry that the pseudoscalar correlator may be 
contaminated by the zero modes~\cite{bcc00,ddh02} particularly at small volumes, we 
look at $\langle \sum_{\vec{x}} A_4(\vec{x},t) A_4(0)\rangle$ to check if the masses 
calculated are the same as those from the $\langle PP\rangle$ correlator.  We plot the
ratios of the pion masses from the $\langle P\,P\rangle$ correlator and the
$\langle A_4\, A_4\rangle$ correlator in Fig.~\ref{ratio_masses}.  We see that
the ratios are basically unity within errors all the way to the smallest quark
mass.  This suggests that at this volume, i.e.\ $V = (3.2\, {\rm fm})^4$,
there is no detectable contamination due to the zero modes.  However, this is
not so at smaller volumes.  We will show results on the $12^3 \times 28$
lattice and discuss the volume dependence in Sec.~\ref{fv}.

It has been shown that the correlator $\langle PP\rangle - \langle SS\rangle$
can get rid of the zero mode contribution~\cite{bcc00} for both the direct and
cross terms. This is a good procedure for medium quark masses. For small quark masses, 
there is a complication due to the quenched $\eta'$-$\pi$ ghost state contribution in the
$\langle SS\rangle$~\cite{bde02} which has a mass close to two times the pion
mass. We have fitted the $\langle PP\rangle - \langle SS\rangle$ correlator
with the inclusion of the ghost state which has a negative weight. We show in 
Fig. \ref{pp-ss} the ratio of pion masses determined from the
$\langle P\,P\rangle$ and the $\langle P\,P\rangle - \langle S\,S\rangle$ correlators.
We note that the errors from the $\langle P\,P\rangle - \langle S\,S\rangle$ are larger for
$m_{\pi} \le 0.4438(36)$ due to the presence of the ghost state which is close to the
ground state pion mass. It is not clear to what extend the additional
complication due to the ghost state is disentangled from that of the zero modes. 
In any case, we find that the pion masses obtained are consistent with those from the 
$\langle P\,P\rangle$ and $\langle A_4\,A_4\rangle$ correlators. Their central values 
differ by 2.5\% at most. We will come back to this issue in the next section when
we fit the quenched chiral log in $m_{\pi}$.

\section{Chiral log in $m_{\pi}$}  \label{chi_log_mpi}

It is predicted that in the quenched approximation, there are quenched chiral
logs~\cite{sha92,bg92} arising from the hairpin diagrams; the flavor-singlet
pseudoscalar correlator does not yield a full $\eta'$ mass due to the absence
of dynamical fermion loops.  Instead, the would-be $\eta'$ propagator gives a
double pole of the Goldstone boson.  The predicted form for the one-loop
formula in $\chi$PT is the following~\cite{sha97,hsw00}
\begin{equation} \label{chi_log}
m_{\pi}^2  = A m \{1 -\delta [\ln(Am /\Lambda_{\chi}^2 ) +1]\} + A_{\alpha} m^2 [1 + 
2 \ln(Am /\Lambda_{\chi}^2) ]  + B m^2 ,
\end{equation}
where the coefficients $A, A_{\alpha}$ and $B$ are given in terms of the
parameters $\Sigma, f, \alpha_{\Phi}$ and $2\alpha_8 - \alpha_5$ in the
quenched effective chiral lagrangian, i.e.
\begin{eqnarray}
A &=& 2 \Sigma/f^2 , \nonumber \\
A_{\alpha} &=& \frac{\alpha_{\Phi} A^2}{3 (4 \pi f)^2} , \nonumber \\ 
B &=& \frac{(2\alpha_8 - \alpha_5)A^2}{(4 \pi f)^2}, 
\end{eqnarray}
and $\Lambda_{\chi}$ is an unphysical cutoff scale which is usually taken in
$\chi$PT to be in the range of $0.5$--$1.0\,{\rm GeV}$.  The parameter
$\alpha_{\Phi}$ is the singlet coupling in the quenched theory.  The quenched
chiral log parameter $\delta$ is
\begin{equation}  \label{delta}
\delta = \frac{m_0^2}{16 \pi^2 N_f f^2},
\end{equation}
where $m_0 \sim 870\,{\rm MeV}$ from the Veneziano model of the $\eta'$ mass.
From this, one estimates that $\delta = 0.183$.

\begin{figure}[b]
\includegraphics[width=0.8\hsize]{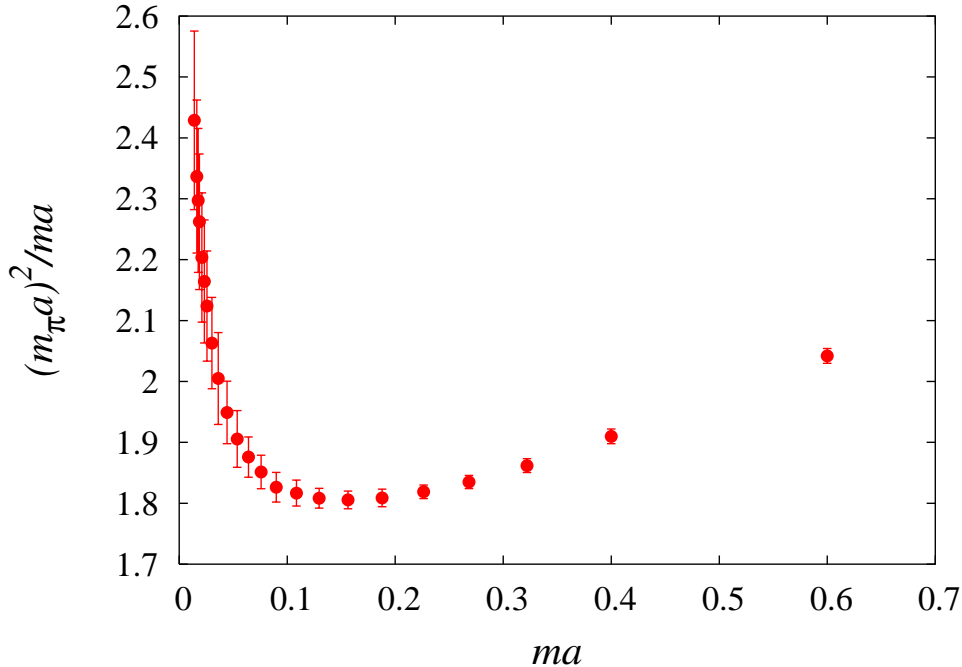}
\caption{\label{mpi_sq_on_m} The ratio $m_{\pi}^2/m$ as a function of the quark
mass $m$.}
\end{figure}

{\small %
\begin{table}[ht]
\caption{\label{summary_delta} Summary of the quenched chiral log calculations
in $m_{\pi}^2$ with different fermion actions and ranges of quark masses.
$m_{\pi, min}$ is the minimum pion mass used in the fitting.  $m_{\pi, trou}$
is the pion mass at the trough of $m_{\pi}^2/m$.  $\mathcal{E}$ is the
percentage of excess of the highest $m_{\pi}^2/m$ at $m_{\pi,min}$ relative to
the corresponding trough at $m_{\pi,trou}$.}
\vspace*{0.2in}
\begin{tabular}{l@{\hspace{0.6em}}l@{\hspace{0.6em}}l@{\hspace{0.6em}}l@{\hspace{0.6em}}l@{\hspace{0.6em}}c@{\hspace{0.6em}}l@{\hspace{0.6em}}l}
\hline
Group                     & Fermion            & $L\,a$  & $m_{\pi, min}$ & $m_{\pi, trou}$ & $\mathcal{E}$ &$\delta$   \\
                          &                    & (fm)    & (MeV)          & (MeV)           &               &           \\
\hline
CP-PACS~\cite{abb00}      & Wilson             & 3.2     & 300  & $\sim$ 550  & $\sim$     8\%   & 0.060(10) -- 0.106(5)\\
Fermilab~\cite{bde00}     & Wilson-Clover      & 2.7     & 290  & $\sim$ 400  & $\sim$     7\%   & 0.073(20)            \\
QCDSF~\cite{qcdsf00}      & Wilson-Clover      & 1.5-2.3 & 400  & $\sim$ 800  & $\sim$    11\%   & 0.14(2)              \\
MILC~\cite{milc01}        & Staggered          & 2.6     & 320  & $\sim$ 700  & $\sim$     8\%   & 0.061(3))            \\
Kim \& Ohta~\cite{ko00}   & Staggered          & 2.6     & 210  & $\sim$ 570  & $\sim$     8\%   & 0.057(13)            \\
Blum et al.~\cite{bcc00}  & Domain Wall        & 3.2     & 390  & $\sim$ 740  & $\sim$    11\%   & 0.07(4)              \\
RCB~\cite{aok01}          & Domain Wall        & 2.4     & 285  & $\sim$ 660  & $\sim$     6\%   & 0.107(38)            \\
BGR~\cite{gat02}          & fixed point        & 2.6     & 210  & $\sim$ 870  & $\sim$ 15-20\%   & 0.17(2)              \\
BGR~\cite{gat02}          & Chirally Improved  & 2.4     & 240  & $\sim$ 930  & $\sim$ 15-20\%   & 0.18(3)              \\
Chiu \& Hsieh~\cite{ch02} & Overlap            & 1.5     & 440  & $\sim$ 770  & $\sim$     11\%  & $\sim$ 0.2           \\
Present work              & Overlap            & 3.2     & 182  & $\sim$ 480  & $\sim$     33\%  & 0.236(33)(37) (one loop) \\
                          &                    &         &      &             &                  & 0.20(3) (cactus) \\
\hline
\end{tabular}
\end{table}
}

There are several calculations to extract the chiral log parameter
$\delta$~\cite{abb00,bde00,qcdsf00,milc01,ko00,bcc00,aok01,gat02,ch02}.  Save
for Refs.~\cite{gat02,ch02}, most of the calculations obtained values for
$\delta$ in the range $0.06 - 0.1$ which are smaller than the expectation.  We
have summarized these results in Table~\ref{summary_delta}.  The present
simulation adopts the overlap fermion action which has exact chiral symmetry on
the lattice, and a highly accurate numerical approach (the matrix sign function
is approximated to better than $10^{-9}$).  This should be a valuable tool to
probe the relevant range of applicability of the one-loop $\chi$PT with pion
mass as low as $\sim 180\,{\rm MeV}$ on a fairly large volume ($3.2\,{\rm
fm}$).  We plot in Fig.~\ref{mpi_sq_on_m} the $\langle PP \rangle$ results for
$m_{\pi}^2/m$ as a function of the bare quark mass $m$.  We see clearly that
below $m = 0.1295$ ($m_{\pi} \sim 480\,{\rm MeV}$), there is a dramatic
divergent behavior.  The ratio of the highest value at $m = 0.014$ ($m_{\pi}
\sim 180\,{\rm MeV}$) to that at the trough at $m = 0.1295$ ($m_{\pi}\sim
480\,{\rm MeV}$) is $\sim 1.33$.  Thus, the percentage excess, $\mathcal{E}$,
of the highest value at the minimum pion mass $m_{\pi,min}$ relative to that at
the trough at $m_{\pi,trou}$ is 33\%.  Most of the calculations studying the
quenched chiral log stopped above $m_{\pi} \sim 300\,{\rm MeV}$; thus, they
seem to be seeing the onset of this behavior with corresponding excess $\sim
6\% - 11\%$.  In Table~\ref{summary_delta}, we summarize the attempts made by
various groups to calculate this quenched chiral log from $m_{\pi}^2/m$ with
different actions and lattices.  Except for the present work, all the results
are fitted to Eq.~(\ref{chi_log}) without the $A_{\alpha}$ term.

We see from Table~\ref{summary_delta} that the results are not consistent with
each other.  There is a large spread in $\delta$ and $m_{\pi, trou}$ which
invokes many issues to ponder: To what extent does the fermion action matter?
Is $\chi$PT valid in the range of pion masses calculated?  If so, is the
one-loop formula sufficient and how does one judge it?  The parameters $A$,
$\delta$, $\Lambda_{\chi}$, and $B$ are not linearly independent in
Eq.~(\ref{chi_log}); how does one fit them?  Practically all the calculations
so far ignored $A_{\alpha}$; does it change the results much if included?  How
large are the finite volume effects for the range of pion masses in the
calculation?  (As will be demonstrated in Section~\ref{fv}, the pion masses
calculated from the $\langle PP\rangle$ correlator in small volumes are
contaminated by the zero modes even up to fairly large quark masses, i.e.\ the
strange quark region.  This could affect the position of $m_{\pi, trou}$ and
the result for $\delta$.)  How does one control the finite volume and zero mode
effects?

Before addressing these questions, we first point out that there is a
fundamental problem in the way the lattice data are usually fitted to the form
predicted by the $\chi$PT\@.  The common practice is to fit the lattice data to
the $\chi$PT result from a high pion mass down to cover the range of calculated
masses and then extrapolate to the physical pion mass.  This can force a fit
but cannot answer the question as to whether the one-loop chiral formula is
applicable for the range of masses considered.  It is possible that the range
of masses will require higher orders in $\chi$PT, or even that the mass range
is outside the radius of convergence of $\chi$PT\@.  Then one does not know the
precise form to fit.  Extrapolating down from the pion mass as high as $1\,{\rm
GeV}$ is simply diametrically opposed to the philosophy of $\chi$PT which is
based on the expansion of small quark masses and momenta.

So, how does one proceed?  To be consistent with the $\chi$PT approach, we
start from the smallest quark mass and ask where the one-loop formula
applies~\cite{note1}.  Since we have come down to the pion mass as low as $\sim
180\,{\rm MeV}$, we will assume that it is in the range where one-loop $\chi$PT
is valid.  We fit the one-loop formula with the five smallest pion masses and
then incrementally add higher pion masses one by one.  If the formula is valid
for the range of masses, then the parameters will remain constant, i.e.\ they
are truly the low energy constants that we seek to evaluate.  On the other
hand, if the parameters begin to change as one includes more higher masses, we
take it as an indication that the one-loop formula starts to lose its
applicability.  This could be due to the fact that higher orders are needed or
that the mass is simply outside the range of validity of $\chi$PT\@.

\begin{table}[ht]
\begin{center}
\caption{\label{t_delta} The low-energy chiral parameters $C_{1}$ and
$C_{1\,L}$ (and $C_{2}$ and $C_{2L}$) are fitted from Eq.~(\ref{eq:C-ind}) with
a maximum of 17 quark masses in correlated and constrained fits for several
ranges of quark masses with the minimum quark mass at 0.0140 (which corresponds
to the smallest pion mass at $182(8)\,{\rm MeV}$).  These are then used to
determine $A$, $\delta$, $A_{\alpha}$, and $B$ from Eq.~(\ref{C_i}).  The
parameters $A$, $\delta$, and $B$ vary with the chiral scale, and are evaluated
at $\Lambda_{\chi}=0.8$.}
\vspace*{0.2in}
\begin{tabular}{c@{\hspace{0.9em}}c@{\hspace{0.9em}}c@{\hspace{0.9em}}c@{\hspace{0.9em}}c@{\hspace{0.9em}}c@{\hspace{0.9em}}c@{\hspace{0.9em}}c@{\hspace{0.9em}}c}
\hline
$m_{max}$& $m_{\pi, max}$  & $C_{1}$    &$C_{1\,L}$     &A              &$\delta$       &$A_{\alpha}$   &$B$            &$\chi^2$/dof\\
\hline
0.02100 &0.2151(61)     &0.82(17)       &-0.368(64)     &1.51(27)       &0.244(40)      &--             &--             &0.48 \\
                                                                                  
0.02333 &0.2244(59)     &0.84(16)       &-0.362(52)     &1.51(24)       &0.239(36)      &--             &--             &0.49 \\
                                                                                  
0.02567 &0.2332(56)     &0.83(15)       &-0.353(51)     &1.48(23)       &0.238(35)      &--             &--             &0.46 \\
                                                                                  
0.03033 &0.2501(54)     &0.84(14)       &-0.350(56)     &1.48(23)       &0.236(35)      &--             &--             &0.44 \\
                                                                                  
0.03617 &0.2693(48)     &0.85(15)       &-0.349(50)     &1.49(22)       &0.234(33)      &--             &--             &0.42 \\
                                                                                  
0.04433 &0.2940(42)     &0.87(14)       &-0.340(50)     &1.50(22)       &0.226(31)      &--             &--             &0.46 \\
                                                                                  
0.05367 &0.3200(43)     &0.87(14)       &-0.315(47)     &1.44(21)       &0.219(32)      &--             &--             &0.58 \\
                                                                                  
0.06417 &0.3470(40)     &1.05(16)       &-0.256(50)     &1.53(22)       &0.167(30)      &--             &--             &0.61 \\
                                                                                  
0.07583 &0.3748(38)     &1.08(16)       &-0.246(48)     &1.55(21)       &0.159(29)      &--             &--             &0.64 \\
                                                                                  
0.08983 &0.4050(38)     &1.13(16)       &-0.233(42)     &1.57(21)       &0.148(26)      &-0.003(167)    &1.55(55)       &0.82 \\
                                                                                                                 
0.10850 &0.4438(36)     &1.16(14)       &-0.227(42)     &1.59(19)       &0.143(24)      &-0.053(132)    &1.39(45)       &0.93 \\
                                                                                                                 
0.12950 &0.4840(32)     &1.14(13)       &-0.228(39)     &1.58(18)       &0.145(22)      &-0.092(125)    &1.42(40)       &0.92 \\
                                                                                                                 
0.15633 &0.5312(31)     &1.16(12)       &-0.228(39)     &1.60(17)       &0.143(21)      &0.003(93)      &1.42(31)       &0.79 \\
                                                                                                                 
0.18783 &0.5829(28)     &1.15(9)        &-0.230(38)     &1.59(13)       &0.145(19)      &-0.001(63)     &1.43(24)       &0.79 \\
                                                                                                                 
0.22633 &0.6416(25)     &1.15(8)        &-0.221(35)     &1.57(12)       &0.141(17)      &-0.093(49)     &1.47(18)       &0.81 \\
                                                                                                                 
0.26833 &0.7017(24)     &1.16(8)        &-0.221(31)     &1.58(11)       &0.140(16)      &-0.046(38)     &1.44(15)       &0.95 \\
                                                                                                                 
0.32200 &0.7743(22)     &1.17(8)        &-0.214(30)     &1.58(11)       &0.135(15)      &-0.070(40)     &1.43(14)       &0.95 \\
                                                                                                                 
0.40000 &0.8740(30)     &1.17(7)        &-0.213(30)     &1.58(11)       &0.135(15)      &-0.079(37)     &1.43(13)       &0.92 \\
\hline
\end{tabular}
\end{center}
\end{table}

\begin{figure}[ht]
\includegraphics[width=0.8\hsize]{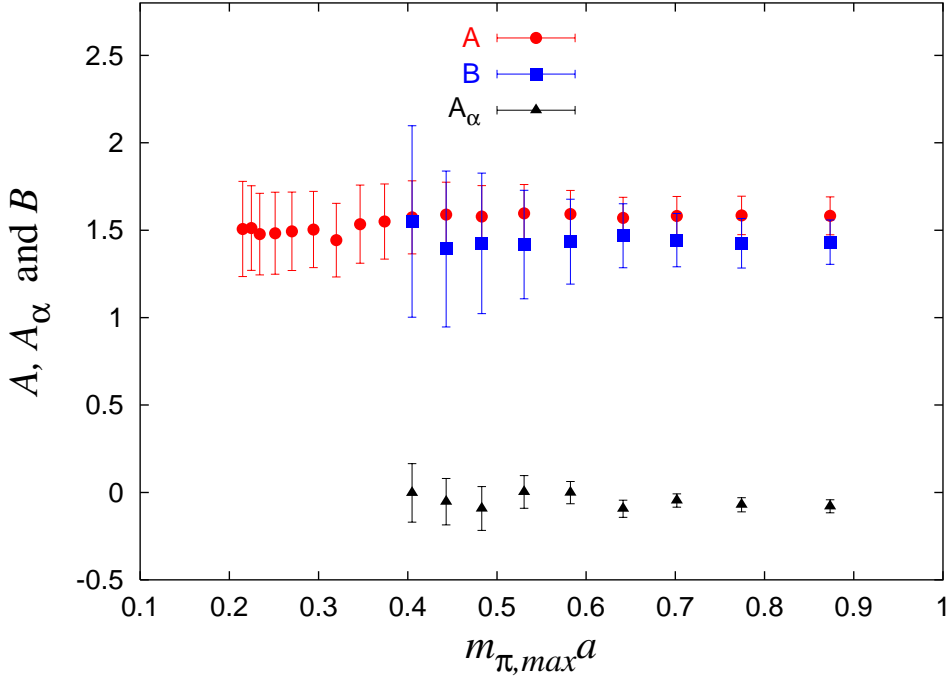}
\caption{\label{A} Fitted $A, A_{\alpha}$, and $B$ as a function of $m_{\pi,
max}$, the maximum pion mass of the fit, for $\Lambda_{\chi} =0.8$.  The
minimum pion mass is at 0.1844(71) ($182(8)\,{\rm MeV}$).}
\end{figure}

\begin{figure}[hb]
\includegraphics[width=0.8\hsize]{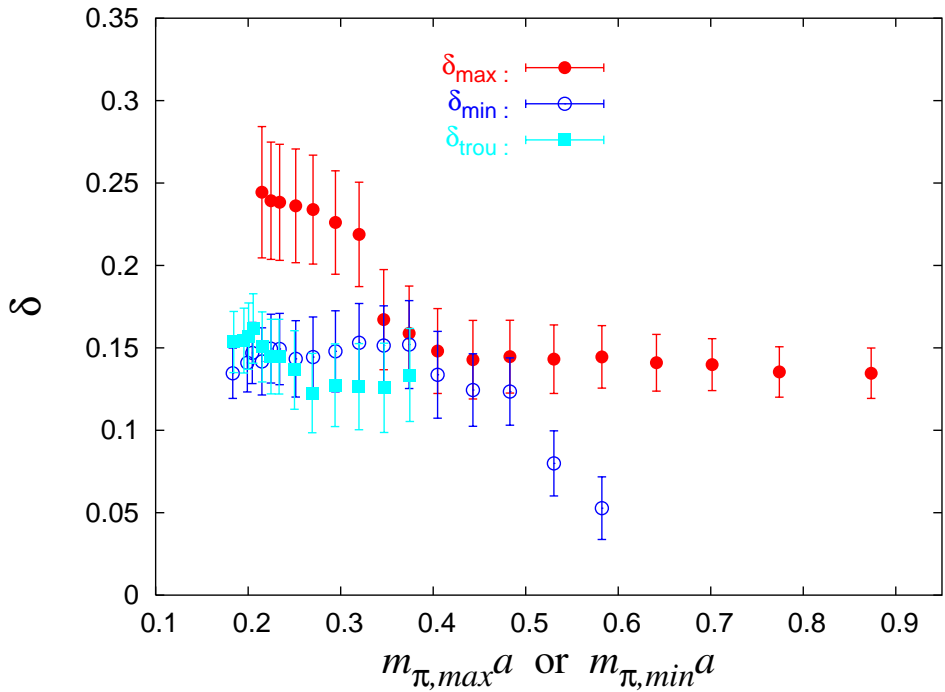}
\caption{\label{quen_delta} The filled circles are the quenched chiral $\delta$
plotted as a function of $m_{\pi, max}$, the maximum pion mass of the fitting
range, with the minimum pion mass fixed at 0.1844(71) ($182(8)\,{\rm MeV}$).
The open circles (filled squares) are those fitted downward from a maximum pion
mass fixed at $0.8740(30)$ ($862(13)\,{\rm MeV}$) (0.6414(21) ($633(10)\,{\rm
MeV}$)) and plotted as a function of $m_{\pi, min}$, the minimum pion mass of
the fitting range.  These results are at $\Lambda_{\chi} =0.8$.}
\end{figure}

\begin{figure}[ht]
\includegraphics[width=0.8\hsize]{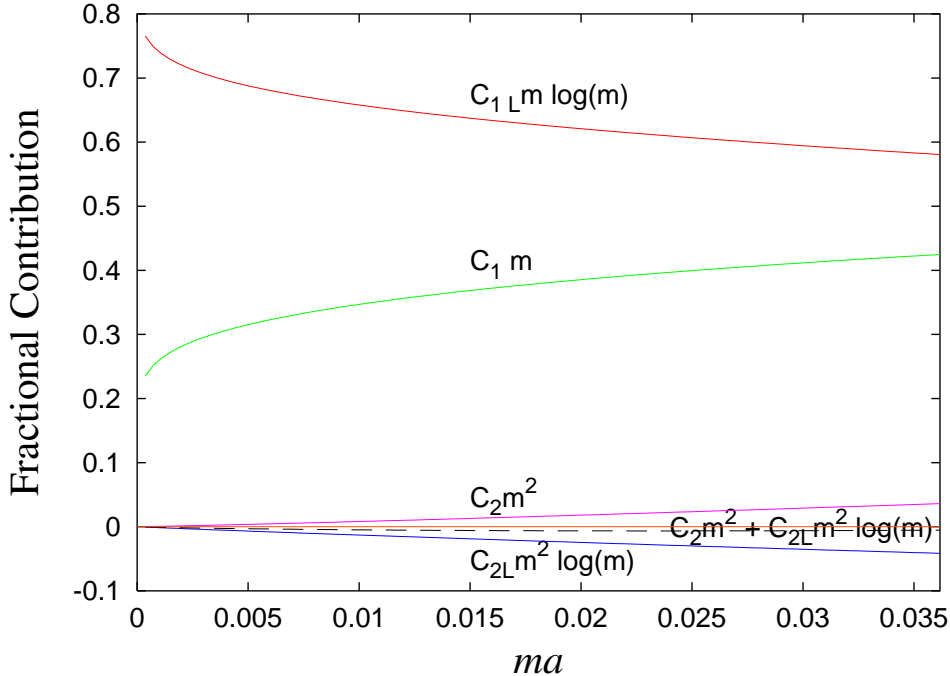}
\caption{\label{C-contri} Fractional contributions for different terms in
Eq.~(\ref{eq:C-ind}).  The input of the $C$ parameters are from the fit with
$m_{max} = 0.03617$. Note the combined contribution $C_1 m^2 + C_{1\,L} m^2 \ln
(m)$ denoted by the dashed line is very small.}
\end{figure}

We mentioned earlier that in the fitting of the one-loop formula, the
parameters $A, \delta$, $\Lambda_{\chi}$, and $B$ are not linearly independent.
We can rewrite Eq. (\ref{chi_log}) as
\begin{equation}   \label{eq:C-ind}
  m_{\pi}^2 = C_{1} m + C_{1\,L} m \ln (m) + C_{2} m^2 + C_{2\,L} m^2 \ln (m)
\end{equation}
where
\begin{eqnarray} \label{C_i}
  C_{1}    &=& A[ 1 - \delta(\ln A - 2 \ln \Lambda_{\chi} +1)], \nonumber \\
  C_{1\,L} &=& - A \delta, \nonumber \\
  C_{2}    &=& A_{\alpha}(1 + 2 \ln A - 4 \ln \Lambda_{\chi}) + B,\nonumber \\
  C_{2\,L} &=& 2 A_{\alpha}, 
\end{eqnarray}
are the independent parameters. Therefore, we have one redundant variable which
we take to be the cutoff scale $\Lambda_{\chi}$.  We adopt a Bayesian-based
constrained curve-fitting method~\cite{lcd01,mor01} with an adaptive procedure
to obtain priors sequentially according to the relative importance of the
parameters.  The details of the constrained curve-fitting algorithm for the
hadron masses will be given elsewhere~\cite{ddh03}.  In each of the quark mass
ranges, the priors for $C_{1}$, $C_{1\,L}$, $C_{2}$, and $C_{2\,L}$ are
obtained from the corresponding $\langle A_4 A_4 \rangle$ correlators.  After
the $C$ parameters are determined from the fit, we solve Eq.~(\ref{C_i}) to
obtain $A$, $\delta$, $A_{\alpha}$, and $B$ for a given $\Lambda_{\chi}$.  We
list their results evaluated at $\Lambda_{\chi} = 0.8$ together with $C_{1}$
and $C_{1\,L}$ in Table~\ref{t_delta}.  We also plot $A$, $A_{\alpha}$, $B$,
and $\delta$ as a function of the maximum pion mass $m_{\pi, max}$ of the
fitting range in Figs.~\ref{A}, and~\ref{quen_delta}.  We should mention that
due to the limited number of gauge configurations, we are not able to carry out
correlated fits with a reasonable $\chi^2$ for more than 17 quark
masses. Therefore, for longer ranges, we choose a maximum of 17 representative
masses to cover the range of the fit.  For $m_{max}$ less than $0.08983$
($m_{\pi} = 0.4050(38)$ or $400(7)\,{\rm MeV}$), we find that the data are
dominated by $C_{1\,L}$ and $C_{1}$.  The contributions from $C_{2}$ and
$C_{2\,L}$ cancel each other and the combined contribution is negligible
compared to those of $C_{1\,L}$ and $C_{1}$.  This is illustrated in
Fig.~\ref{C-contri} for a typical fit.  As a result, it is practically
impossible to obtain a reliable fit for $C_{2}$ and $C_{2\,L}$.  We decided to
fit these low mass data using a prior for $C_{2\,L}$ which is limited to less
than 2 during the fitting procedure for the $\langle A_4 A_4 \rangle$
correlators.  We tried using the priors without such a limit and also tried
dropping $C_{2}$ and $C_{2\,L}$ all together in these low mass ranges and found
that the fitted $C_{1}$ and $C_{1\,L}$ are not changed within errors.  The
central values for $\delta$ are changed only by $1 - 2$\%.  In view of this
inability to fit $C_2$ and $C_{2\,L}$, we do not quote the fitted $A_{\alpha}$
and $B$ below $m_{max} = 0.08983$ ($m_{\pi}=0.4050(38)$) where the errors are
larger than their central values.  Hopefully, with better statistics, one will
be able to extract the parameters $A_{\alpha}$ and $B$ in the range $m_{\pi} <
400(7)\,{\rm MeV}$.

We see in Fig.~\ref{A} that values for $A$ are fairly stable throughout the
range of the fit.  $A_{\alpha}$ and $B$ are also stable in the range where
$m_{max} \ge 0.08983$ and $A_{\alpha}$ is consistent with zero in this range.
On the contrary, $\delta$ in Fig.~\ref{quen_delta} is constant up to $m_{max}
\sim 0.044$ which corresponds to $m_{\pi, max} \sim 300\,{\rm MeV}$.  Beyond
that, it has a sharp drop.  We take the fact that $\delta$ is sensitive to the
range of masses as an indication that the one-loop chiral formula in
Eq.~(\ref{chi_log}) is valid {\it only\/} up to $m_{\pi} \sim 300\,{\rm MeV}$.
Below this value, the parameters in the formula do not exhibit mass dependence
and thus they are the low energy constants of the effective chiral theory at a
certain chiral scale (e.g.  $\Lambda_{\chi} = 0.8$).

Next, we simulate what has been commonly done in lattice calculations.  We fix
the maximum pion mass at $m_{\pi} = 0.8740(30)$ ($862(13)\,{\rm MeV}$) and
incrementally include lower pion masses to fit Eq.~(\ref{eq:C-ind}).  We find
that no matter how many low pion masses are included, the results are not
compatible with those starting from $m_{\pi, min} = 0.1844(71)$ ($182(8)\,{\rm
MeV}$) and ending at $m_{\pi, max} \sim 300\,{\rm MeV}$.  We illustrate this by
plotting, in Fig.~\ref{quen_delta}, the fitted $\delta$ (open circles) as a
function of the minimum pion mass $m_{\pi, min}$ for the range of the fit.  We
see that when pion masses down to $300 - 400\,{\rm MeV}$ are included, the
fitted $\delta$ is in the range of $\sim 0.15$ which is consistent with those
found in Table~\ref{summary_delta} with a similar mass range and can explain
why most of the calculations end up with a small $\delta$.  The only exception
is Ref.~\cite{ch02} which obtains a large $\delta \sim 0.2$ with $m_{\pi, min}
\sim 440\,{\rm MeV}$.  We speculate that this may be due to the small volume
used ($L = 1.5\,{\rm fm}$).  As will be explored further in Section~\ref{fv},
we find that the zero modes make a sizable contribution to pion masses as high
as $800\,{\rm MeV}$ on a lattice with $L = 2.4\,{\rm fm}$, and likely higher at
smaller volumes.  It would be sensible to check if this is the case by
comparing pion masses obtained from the $\langle PP\rangle$ correlator to those
from $\langle A_4 A_4 \rangle$ as will be done in Section~\ref{fv}.

In view of the fact that the rise of $m_{\pi}^2/m$ left of the trough at
$m_{\pi} \sim 480$ MeV is mostly due to the chiral log while the rise right of
the trough is mostly due to the $m^2$ term, we first fit the points around the
trough (i.e. from $m = 0.07583$ to 0.22633) with the form in
Eq. (\ref{eq:C-ind}) but without the $A_{\alpha}$ term and then extend the
range by including smaller quark masses. The results of $A, \delta$ and $B$
turn out to be very close to the ones starting from the maximum pion mass at
$m_{\pi} = 0.8740(30)$ ($862(13)\,{\rm MeV}$). We show the results of $\delta$
(labeled $\delta_{trou}$) as a function of the minimum pion mass $m_{\pi, min}$
in Fig.~\ref{quen_delta} and find them almost the same as $\delta_{min}$ at
small quark masses and similarly smaller than those where the pion masses are
restricted to lower than $\sim 300\,{\rm MeV}$.
 
\begin{figure}[ht]
\includegraphics[width=0.8\hsize]{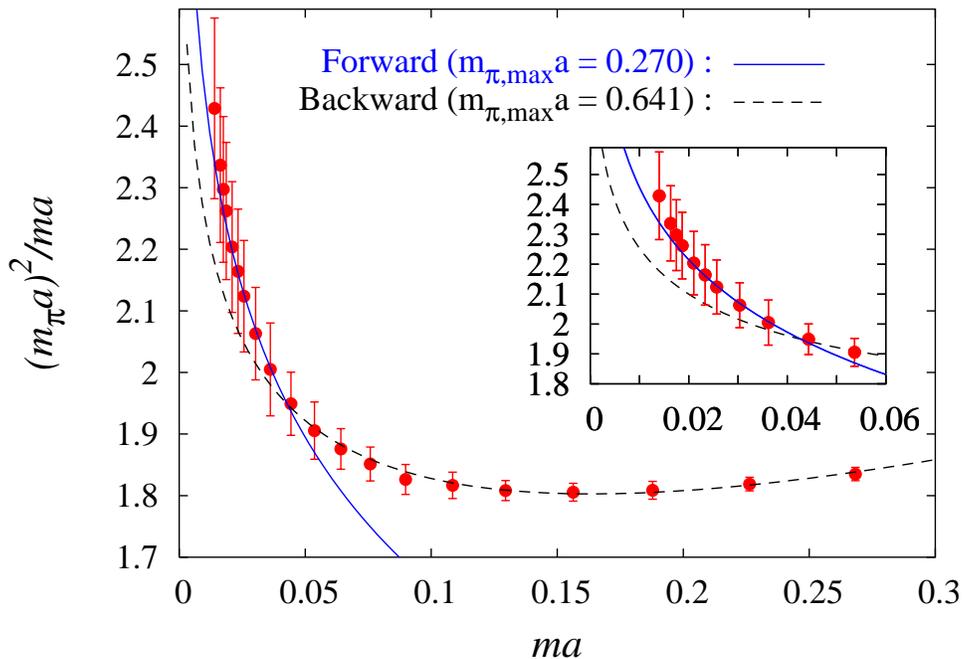}
\caption{\label{m_pi_fit} The fit to $m_{\pi}^2/m$ as a function of the
quark mass $m$. The solid line is the forward fitting ending at $m_{\pi} =
0.269(5)$ ($m = 0.03617$). The dash line is the backward fitting from $m_{\pi}
= 0.641(2)$ ($m = 0.22633$) to the smallest quark mass.  The insert is a blowup
for the small quark mass region.}
\end{figure}

For comparison, we show in Fig.~\ref{m_pi_fit} the fit of $m_{\pi}^2/m$ for the
forward fitting ending at $m_{\pi} = 0.269(5)$ ($m = 0.03617$) (solid line) and
the backward fitting from $m_{\pi} = 0.641(2)$ ($m = 0.22633$) to the smallest
quark mass (dash line).

\begin{figure}[hb]
\includegraphics[width=0.7\hsize]{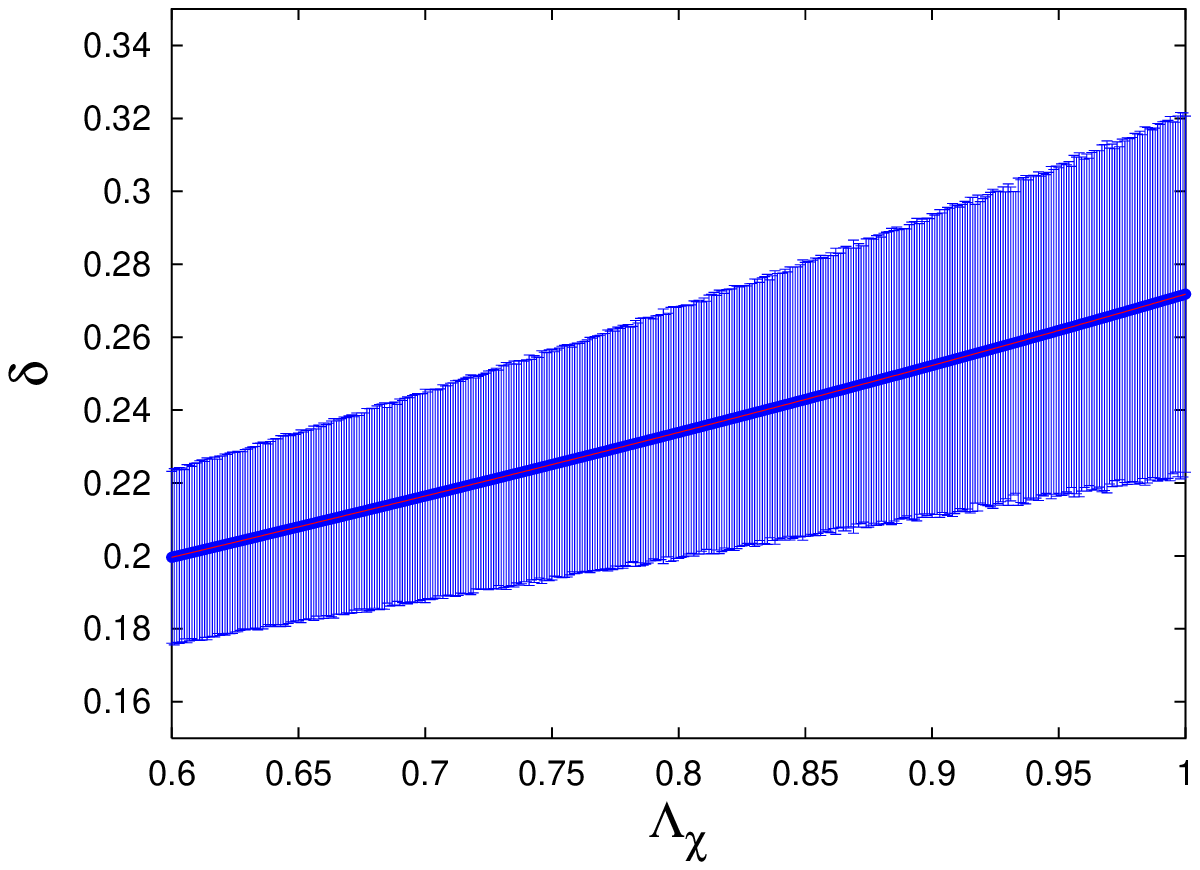}\\
\includegraphics[width=0.7\hsize]{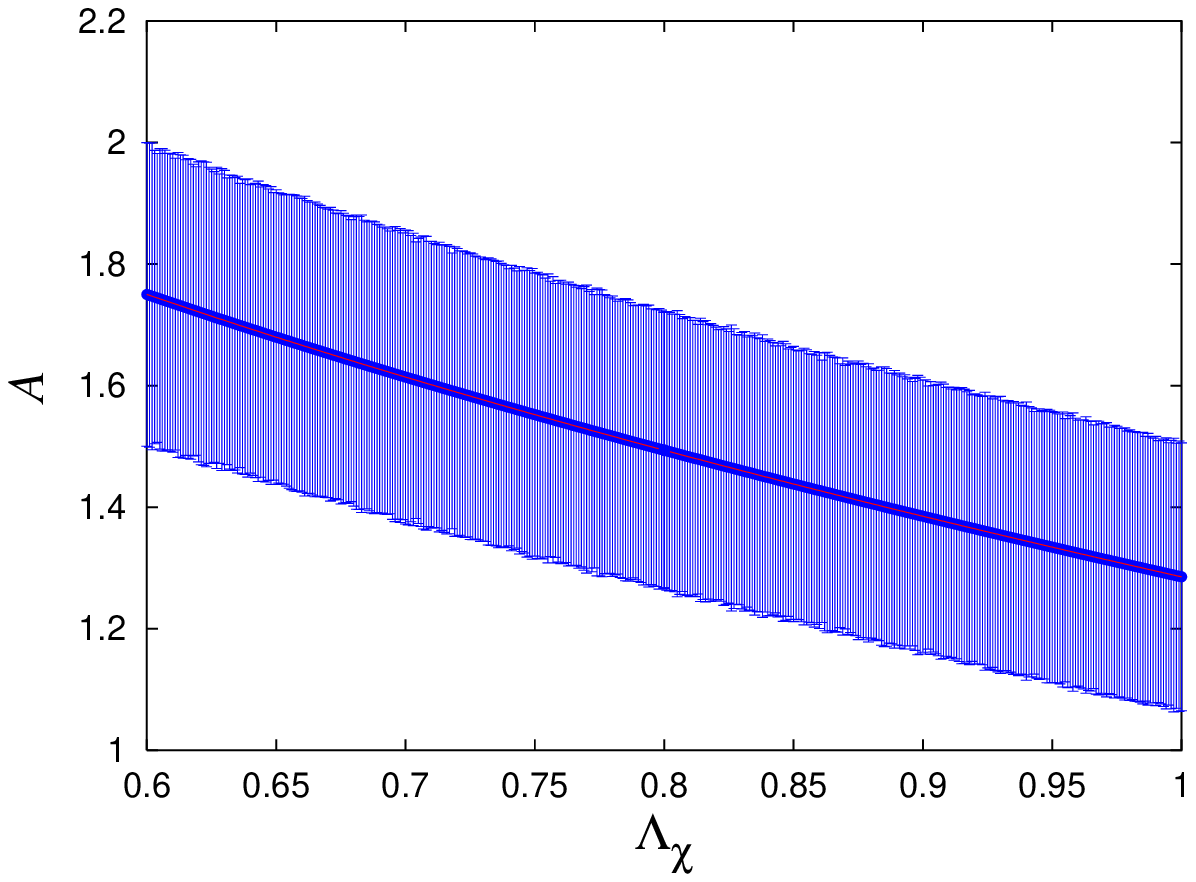}
\vspace*{-0.1in}
\caption{\label{running}
The running of $\delta$ and $A$ and as a function of $\Lambda_{\chi}$ according
to the one-loop formula.  The input of $C_{1}$ and $C_{1\,L}$ are for the fit
with $m_{max} = 0.03617$.}
\end{figure}

As one can see from Eq.~(\ref{C_i}), the parameters $A$, $\delta$, and $B$
depend logarithmically on $\Lambda_{\chi}$.  In Fig.~\ref{running} we show how
$A$ and $\delta$ vary as a function of $\Lambda_{\chi}$.  The $C_{1}$ and
$C_{1\,L}$ are taken from the fit for $m_{max} = 0.03617$ ($m_{\pi} =
0.2693(48)$ or $266(6)\,{\rm MeV}$) in Table~\ref{t_delta}.  Incidentally, for
a given $C_{1}$ and $C_{1\,L}$, there is a maximal value
\begin{equation}
\Lambda_{\chi,{\rm max}}= \sqrt{C_{1\,L}}\,\, e^{C_{1}/2C_{1\,L}}
\end{equation}
(for which $\delta = 1$ and $A=C_{1}$) beyond which no solution exists.  But
this is well beyond the range of applicability of Eq.~(\ref{chi_log}).  We
should point out that the fact $A$ runs with $\ln \Lambda_{\chi}$ is a peculiar
feature of quenched QCD\@.  This is not the case in full QCD where $A$ is
basically the quark condensate where there is no quenched chiral
$\delta$~\cite{sha92}.  From Eq.~(\ref{C_i}), we see that the combination $
A\delta$ is independent of the chiral scale.  This makes $\delta$ also run with
$\ln \Lambda_{\chi}$.  This is not physical, but rather is due to the fact that
Eq.~(\ref{chi_log}) contains only the leading term in the expansion of
$\delta$~\cite{sharpe03}.  To see this, consider the example where the leading
logarithm is re-summed for the ``cactus diagrams''~\cite{sha92} (one is not
concerned with $A_{\alpha}$ and $B$ here) which leads to
\begin{equation}
m_{\pi}^2 = m^{\frac{1}{1 + \delta}}(A \Lambda_{\chi}^{2\delta})^{\frac{1}{1 + \delta}}.
\end{equation}
We see that the $m$ and $\Lambda_{\chi}$ dependence of $\delta$ are separable
which implies that $\delta$ does not run with $\Lambda_{\chi}$, and the
unphysical $\Lambda_{\chi}$ can be absorbed in $A$ with
\begin{equation}
A \propto \Lambda_{\chi}^{- 2\delta}
\end{equation}
which gives
\begin{equation}
\frac{d\,\ln A}{d\,\ln \Lambda_{\chi}^2} = - \delta.
\end{equation} 
This is the same as obtained from Eq.~(\ref{chi_log}) when one requires that
$m_{\pi}^2$ be independent of $\Lambda_{\chi}$ and assumes that $\delta$ does
not depend on $\Lambda_{\chi}$.

We shall report our final results based on the fit of pion mass from $0.184(7)$
($m= 0.014$) to $0.269(5)$ ($m = 0.03617$) with 9 quark masses. This choice is
based on the observation that $\delta$ begins to decrease beyond $m_{max} =
0.03617$ ($m_{\pi\,max} = 0.269(5)$). Although this happens to coincide with
the minimum in $\chi^2/dof$ in Table \ref{t_delta}, we should stress that a
minimal $\chi^2/dof$ is not and should not be the criterion for such a
choice. As we pointed out earlier, the primary difficulty in chiral
extrapolation is the uncertainty about the valid range of one-loop formula in
$\chi$PT. A $\chi^2/dof$ less than unity does not imply that the fitting
formula is correct. One way to judge is the criterion that we adopt,
namely the one-loop formula is not valid when the fitted parameters do not
remain constant beyond certain $m_{\pi\,max}$.

As we said earlier, we are not able to obtain a reliable fit for $A_{\alpha}$,
and $B$ at these low masses.  As for the finite volume systematic errors, we
note that the finite volume correction for the lowest $m_{\pi}^2$ is estimated
to be 5.4\% and the second lowest $m_{\pi}^2$ is estimated to be 3.6\%.  (This
will be explained in more detail in Section~\ref{fv}.)  We thus estimate the
systematic errors due to the finite volume to be at the few percent level which
are smaller than our current statistical errors.  We also dropped the last
quark mass and fitted the 8 quark masses from $m = 0.01633$ ($m_{\pi} =
0.195(7)$) to $m = 0.03617$ ($m_{\pi} = 0.269(5)$) and found that $A$ and
$\delta$ change only about 1\%, which are much smaller than the statistical errors.

We quote our final results for the low-energy parameters of the quenched 
one-loop chiral perturbation theory at $\Lambda_{\chi} = 0.8\, {\rm GeV}$ close to 
the $\rho$ mass in Table~\ref{low_energy_p}.  In view of the unphysical dependence 
of $\delta$ on $\Lambda_{\chi}$ at the the order specified in Eq.~(\ref{chi_log}), 
we shall estimate the systematic error to be the average deviation from the value at
$0.8\,{\rm GeV}$ of the values at $\Lambda_{\chi} = 1\,{\rm GeV}$ and
$0.6\,{\rm GeV}$.  We see that the central value of $\delta$ is somewhat larger
than the phenomenological value of $0.183$ in Eq.~(\ref{delta}), but they are
consistent within errors.

\begin{table}[ht]
\begin{center}
\caption{\label{low_energy_p} The low energy parameters in the quenched chiral
lagrangian are given as a fit of Eq.~(\ref{eq:C-ind}) from $m = 0.014$
($m_{\pi} = 182(8)\,{\rm MeV}$) to $m = 0.03617$ ($m_{\pi} = 266(6)\,{\rm
MeV}$) at $\Lambda_{\chi} = 0.8\,{\rm GeV}$.  The second error in $\delta$ is
the estimated systematic error determined as the average deviation at
$\Lambda_{\chi} = 1$ and $0.6\,{\rm GeV}$ from the value at $0.8\,{\rm GeV}$.}
\vspace*{0.2in}
\begin{tabular}{c@{\hspace{1em}}|@{\hspace{1em}}c@{\hspace{2em}}c}
\hline
$\Lambda_{\chi}$ & $A$ & $\delta$\\
 \hline
0.8 GeV & 1.46(22) GeV & 0.236(33)(37)  \\
 \hline
\end{tabular}
\end{center}
\end{table}

\begin{figure}[ht]
\includegraphics[width=0.8\hsize]{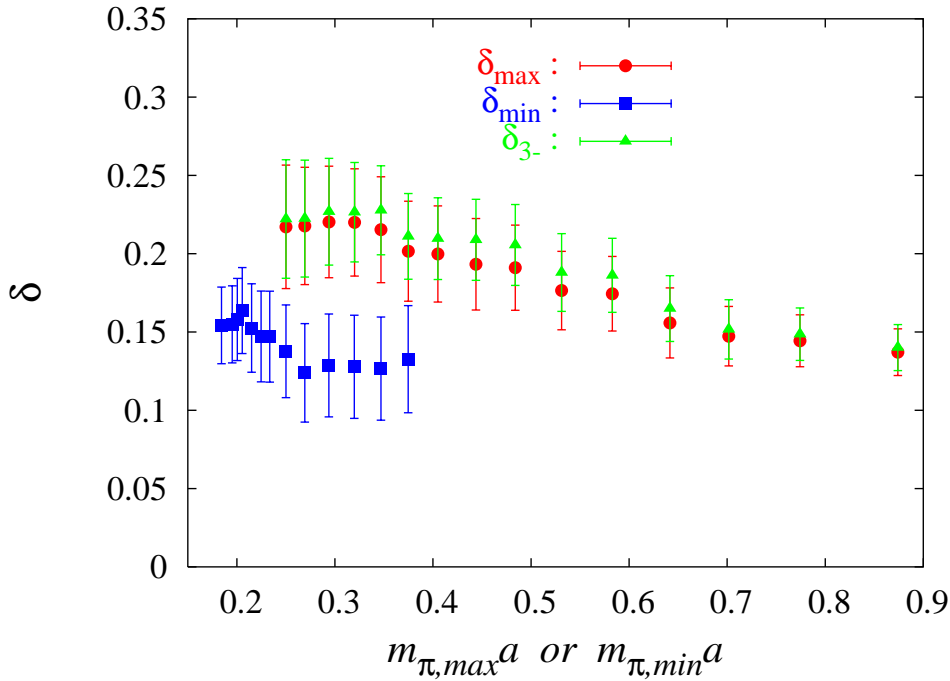}
\caption{\label{delta_cactus} The filled circles are the scale independent
chiral $\delta$ plotted as a function of $m_{\pi, max}$, the maximum pion mass
of the fitting range, with the minimum pion mass fixed at 0.1844(71)
($182(8)\,{\rm MeV}$). The same are for the triangles ($\delta_{3-}$) except
the starting minimum pion mass is at 0.2055(61) ($203(7)\,{\rm MeV}$.  The
filled squares are those fitted downward from a maximum pion mass fixed at
0.6416(25) ($633(10)\,{\rm MeV}$) and plotted as a function of $m_{\pi, min}$,
the minimum pion mass of the fitting range.}
\end{figure}

\begin{figure}[ht]
\includegraphics[width=0.8\hsize]{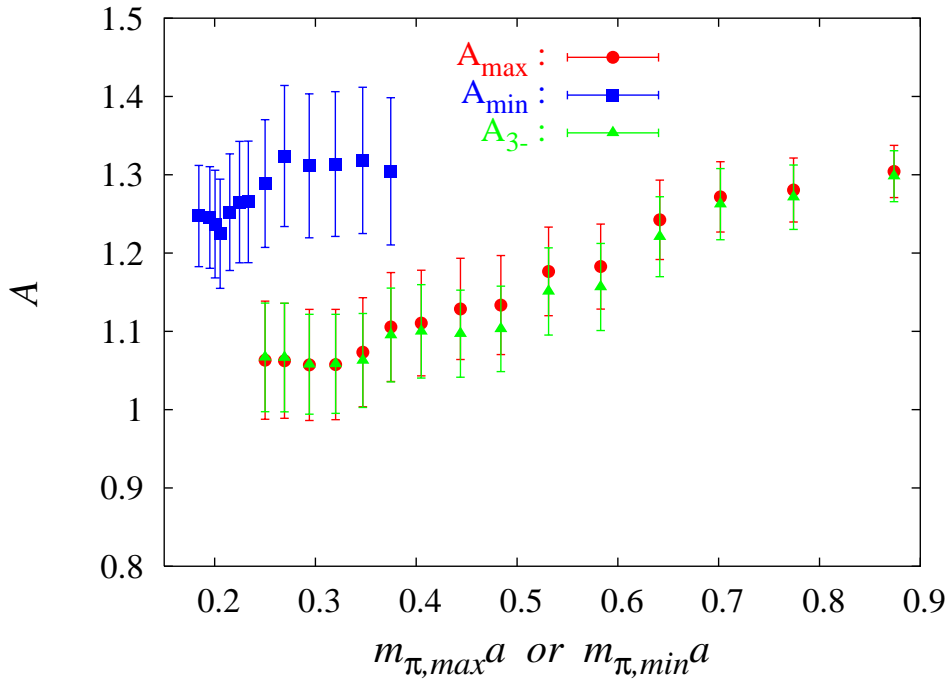}
\caption{\label{A_cactus} The circles/triangles are fitted $A$ as a function of
$m_{\pi, max}$ starting from $m = 0.014/0.01867$.  The squares are those fitted
downward starting from $m = 0.22633$ as a function of $m_{\pi, min}$.}
\end{figure}

\subsection{Fitting of Re-summed Cactus Diagrams}

  Since at small quark masses, the higher order results with $(\delta
\ln(m))^n$ corrections are important, it is pertinent to fit the re-summed
cactus diagram form~\cite{sha92} for the chiral log which leads to a scale independent
$\delta$. Thus, we shall fit with the power form
\begin{equation}  \label{cactus}
m_{\pi}^2 =  A\,m^{\frac{1}{1 + \delta}} + B\,m^2.
\end{equation}

\begin{figure}[ht]
\includegraphics[width=0.8\hsize]{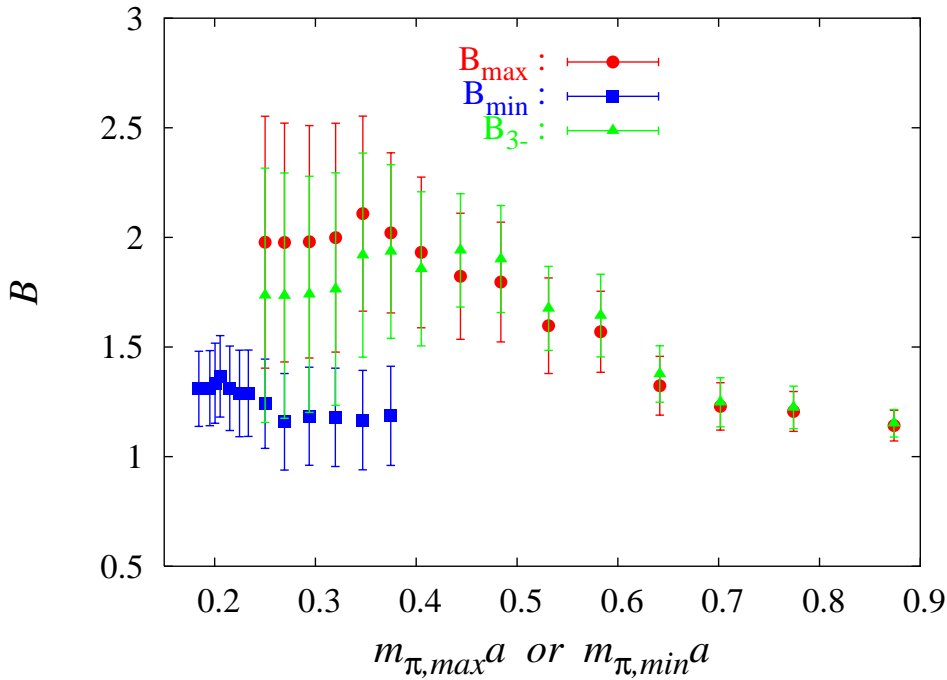}
\caption{\label{B_cactus} The circles/triangles are fitted $B$ as a function of
$m_{\pi, max}$ starting from $m = 0.014/0.01867$).  The squares are those
fitted downward starting from $m = 0.22633$ as a function of $m_{\pi, min}$.}
\end{figure}

Similar to fitting the one-loop formula in Eq. (\ref{eq:C-ind}), we start from
the smallest quark mass and fit upward first. Since at very small quark masses
the $B$ contribution is very small, we put a very weak constraint at $1.8
\pm 1$ which covers the range of $B$ as fitted in one-loop formula in Table~\ref{t_delta}.  
The resultant 
$\delta, A$ and $B$ are presented in Figs. \ref{delta_cactus}, \ref{A_cactus}, and
\ref{B_cactus}. We see that, in contrast to the sharp transition at
$m_{\pi, max} \sim 300\,{\rm MeV}$ in the one-loop formula fit, the transition
for $\delta$ is smoother. The behavior for $A$ and $B$ are similarly changed. This is 
to be expected, since the power form takes into account the re-summed cactus diagrams 
to all orders. 
As such, it should extend the range of applicability of the quenched $\chi$PT to
beyond $m_{\pi} \sim 300\,{\rm MeV}$ for the one-loop case, provided that it
dominates the higher loop corrections. From the fitted results, it appears that
the power form gives a stable fit to the trough of $m_{\pi}^2/m$ and is perhaps
valid to $m_{\pi} \sim 500 - 600\,{\rm MeV}$. Above $m_{\pi} \sim 640\,{\rm
MeV}$, all the parameters ($\delta, A$ and $B$) are different from those below
$m_{\pi, max} \sim 300 - 400\, {\rm MeV}$ by two sigmas. We shall quote the
result of the scale independent $\delta$ at $m = 0.08083 (m_{\pi} = 400(7)\,{\rm MeV})$
where $\delta = 0.20(3)$. We have fitted also the pion mass from the
$\langle P\,P - S\,S\rangle$ correlator and find that the changes of $\delta$ and 
$A$ are small, at a few percent level. In the range between $m_{\pi} = 0.2501(54)$
and 0.4840(32), the change is at the 1\% level. For example, $\delta = 0.20(4)$
at $m = 0.08083 (m_{\pi} = 400(7)\,{\rm MeV})$.

We also fitted the points around the trough ($m = 0.07583$ to 0.22633) and
then incrementally added smaller quark masses in the fit, as we did for the
one-loop formula. The results are plotted in Figs. \ref{delta_cactus},
\ref{A_cactus}, and \ref{B_cactus} as a function of $m_{\pi, min}$ and labeled
$\delta_{min}, A_{min}$ and $B_{min}$. We see again that they do not reproduce
those parameters which are obtained from fitting the range from $m = 0.014$ to
0.08983 ($m_{\pi} = 0.4050(38)$).

\begin{figure}[ht]
\includegraphics[width=0.8\hsize]{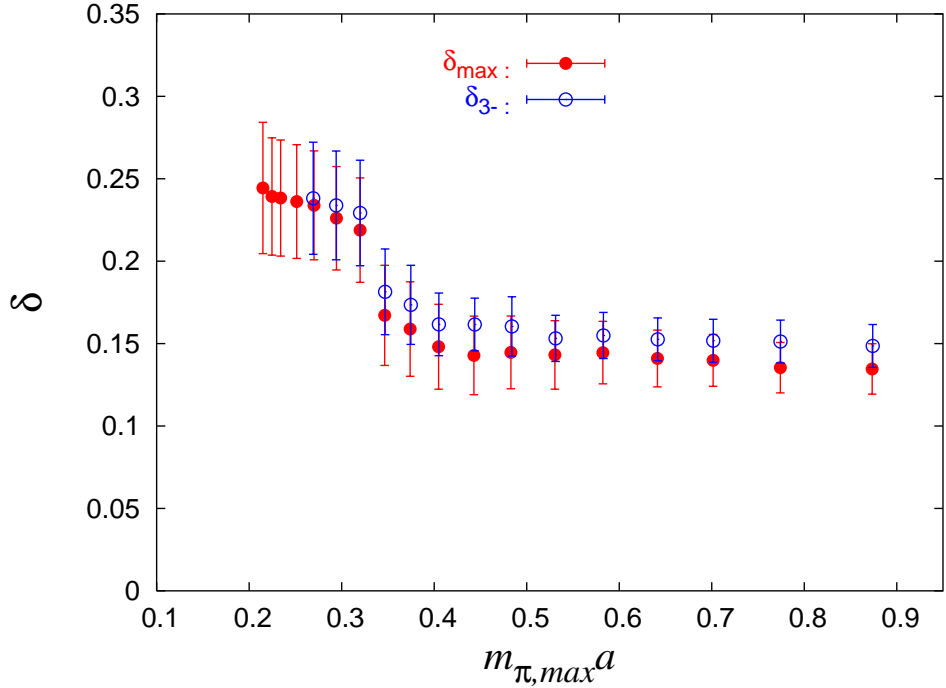}
\caption{\label{delta_log_3} The filled circles are the chiral $\delta$ ($\delta_{max}$) 
plotted as a function of $m_{\pi, max}$, the maximum pion mass
of the fitting range, with the minimum pion mass fixed at 0.1844(71)
($182(8)\,{\rm MeV}$). The same are for the open circles ($\delta_{3-}$) except
the starting minimum pion mass is at 0.2055(61) ($203(7)\,{\rm MeV}$.}
\end{figure}

As we are concerned about the finite volume effect (this will be addressed in
Sec.~\ref{fv}) for small quark masses, we drop the three smallest quark masses and
then fit the pion mass upward. We plot the results in
Figs. \ref{delta_cactus}, \ref{A_cactus}, and \ref{B_cactus} as a function of
$m_{\pi, max}$ and labeled $\delta_{3-}, A_{3-}$ and $B_{3-}$.  We see that in
most cases they are only a few percent different from those ($\delta_{max},
A_{max}$ and $B_{max}$) without dropping the last three quark masses. We have
also performed the same test for the one-loop log formula fit and find that the
parameters are also not changed much. We plot in Fig. \ref{delta_log_3} the results  
of $\delta_{3-}$ from this fit together with $\delta_{max}$ as obatined from the
one-loop log fit without dropping the last 3 points. We see that the characteristic
transtion around $m_{\pi} \sim 300$ MeV is still visible.

\section{Finite volume effects}  \label{fv}

\begin{figure}[bh]
\includegraphics[width=0.8\hsize]{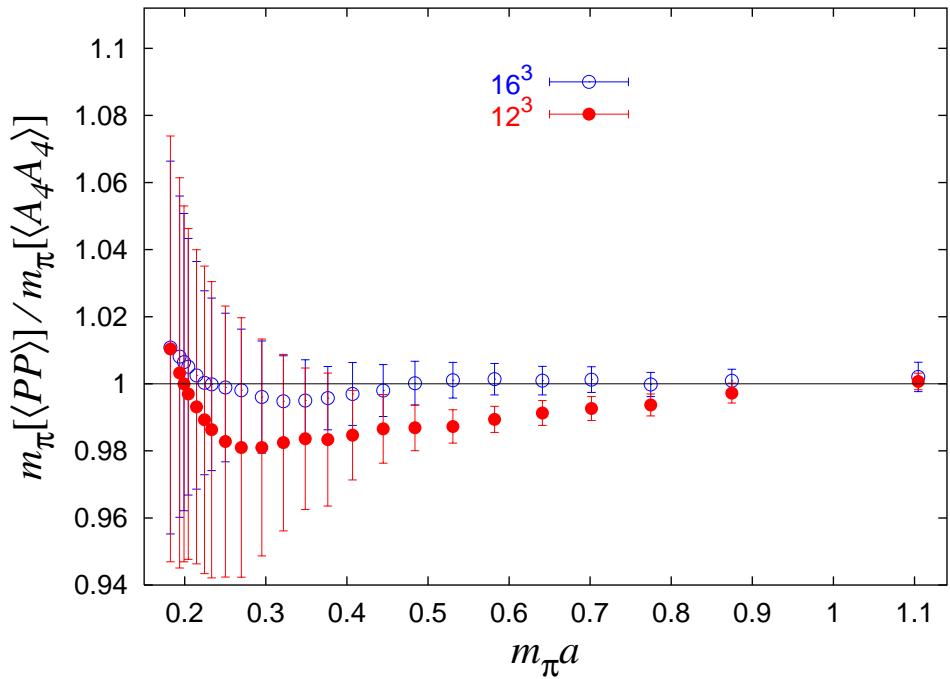}
\caption{\label{r_P_A4} Ratio of pion masses as calculated from the $\langle
P\,P\rangle$ and $\langle A_4\,A_4\rangle$ correlators as a function of
$m_{\pi}$ from the $\langle A_4\,A_4\rangle$ correlator for both the $16^3
\times 28$ and $12^3 \times 28$ lattices.}
\end{figure}

Intertwined in the discussion of the validity of the one-loop $\chi$PT for a
certain mass range are the finite volume effects.  There are two questions to
consider: First of all, how does one know how large the zero mode contributions
are, and how much do they affect the pion mass in the range of the fitting?
Secondly, how large is the finite volume effect for the pion mass where the
finite box is not larger than 4 times the Compton wavelength of the pion? We
considered the zero mode contribution in Section~\ref{pion_zero} where we
examined the ratio of pion masses from the $\langle P\,P\rangle$ and $\langle
A_4\,A_4\rangle$ correlators for the $16^3 \times 28$ lattice and found that
the ratio is basically unity within errors for the range of quark masses that
we are concerned with.  We now look at the same ratio in a smaller lattice
$12^3 \times 28$ ($La = 2.4\,{\rm fm}$ in this case).  They are plotted in
Fig.~\ref{r_P_A4} together with those from the $16^3 \times 28$ lattice.  We
see that the ratio for the $12^3 \times 28$ lattice deviates from unity by two
sigmas for pion mass in the range of 0.5 to 0.8, or $\sim 500$ to $800\,{\rm
MeV}$.  The central value of the ratio goes down to 98\% at $m_{\pi} \sim
300\,{\rm MeV}$ before it turns back to around unity for small pion mass due to
the finite volume effect.  This is a clear indication that the zero mode
contribution to the $\langle P\,P\rangle$ propagator is visible even for pion
mass as heavy as $\sim 800\,{\rm MeV}$.  Most of the calculations compiled in
Table~\ref{summary_delta} have volumes similar to our $12^3 \times 28$ lattice,
therefore one needs to be concerned about the zero mode contamination which is
a finite volume artifact.  Since the zero mode contribution varies as
$\frac{1}{\sqrt{V}}$~\cite{bcc00,ddh02}, the effect of the zero mode
contribution will be larger for calculations with still smaller lattice sizes.
It would be useful to compare the pion masses from $\langle P\,P\rangle$ and
$\langle A_4\,A_4\rangle$ correlators to see how different they are in these
cases.

\begin{figure}[t]
\includegraphics[width=0.8\hsize]{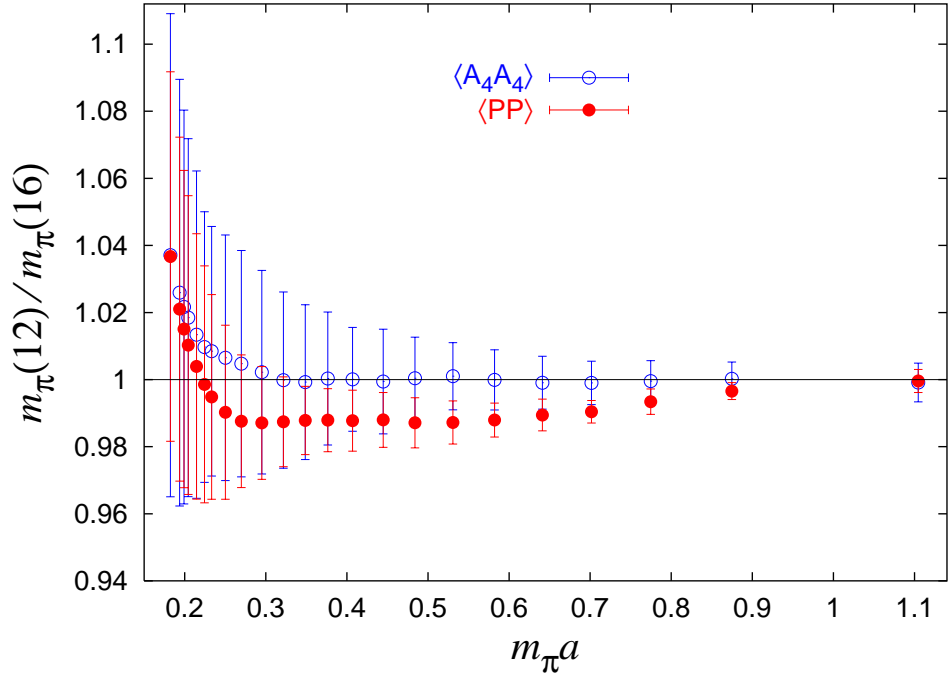}
\caption{\label{m_pi_12_16_ratio} Ratio of pion masses between the $12^2 \times
28$ and $16^3 \times 28$ lattices for both the $\langle A_4\,A_4\rangle$ and
the $\langle P\,P\rangle$ correlators.}
\end{figure}

To check the finite volume effect, we consider the pion mass from the $\langle
A_4\,A_4\rangle$ correlator which is less affacted by the zero mode contamination.  
We plot the ratio of pion masses between the $12^3 \times 28$ and $16^3 \times 28$
lattices in Fig.~\ref{m_pi_12_16_ratio} as a function of $m_{\pi}$.  We see
that the ratio is basically unity for pion mass greater than $\sim 300\,{\rm
MeV}$.  Below this, the pion mass from the smaller volume, i.e.\ the $12^3
\times 28$ lattice, becomes higher.  The worst deviation, at the smallest pion
mass of $182(8)\,{\rm MeV}$, is $3.7\%$.  For comparison, we also plot in
Fig.~\ref{m_pi_12_16_ratio} the corresponding ratio of $m_{\pi}$ from the
$\langle P\,P\rangle$ correlator.  Contrary to the $\langle A_4\,A_4\rangle$
correlator, we see that the ratio has a dip between the $m_{\pi}$ range of 250
and $800\,{\rm MeV}$, again reflecting the fact that the pion mass calculated
from the $\langle P\,P\rangle$ correlator on the $12^3 \times 28$ lattice has
observable zero mode contributions.

To assess the finite volume correction, we first consider the leading finite
volume correction calculated by replacing the infinite volume meson propagator
with the finite volume counterpart.  The correction~\cite{bg92} for $m_{\pi}^2$
is
\begin{equation}
\Delta (m_{\pi}^{1-loop})^2 = \frac{m_{\pi}^2}{4 \pi^2 f^2}(\mu^2 - \alpha m_{\pi}^2)
\sqrt{\frac{2 \pi}{m_{\pi}L}} e^{-m_{\pi}L},
\end{equation}
where $\mu^2 = m_{\eta'}^2 - m_{\eta}^2/2 - m_{\pi^0}^2/2 = (871\, {\rm
MeV})^2$.  With $\alpha m_{\pi}^2 \ll \mu^2$~\cite{bg92}, we can estimate the
percentage difference of $m_{\pi}^2$ between the $12^3 \times 28$ and the $16^3
\times 28$ lattices for the smallest $m_{\pi} = 182(8)\,{\rm MeV}$ to be
\begin{equation}
\frac{\Delta (m_{\pi}^{1-loop})^2 (12) - \Delta (m_{\pi}^{1-loop})^2 (16)}{m_{\pi}^2}
= 0.13\%.
\end{equation}
This is much smaller than the Monte Carlo calculation of 7.4\% derived from the
$\langle A_4\,A_4\rangle$ correlators in Fig.~\ref{m_pi_12_16_ratio}.  Next, we
use the empirical $L^3$ dependence~\cite{fuk92} to estimate the finite volume
correction with
\begin{equation}
m_{\pi}(L) = m_{\pi}(L = \infty) + \frac{a}{L^3}.
\end{equation}
For our lowest $m_{\pi}(L = 16) = 182(8)\,{\rm MeV}$ where the corresponding
$m_{\pi} (L = 12)$ is $3.7\%$ higher, we obtain the finite volume correction
for the $m_{\pi}(L = 16) = 182(8)\,{\rm MeV}$ to be $2.7\%$ and the correction
for $m_{\pi}^2(L = 16) = (182(8) {\rm MeV})^2$ to be $5.4\%$.  Similarly, for
the second lowest pion mass at $193(7)\,{\rm MeV}$, the correction is $3.6\%$
for $m_{\pi}^2(L = 16)$.  These are smaller than the statistical errors for the
fitted low-energy parameters in Section~\ref{chi_log_mpi}.

We are not in a position to fully address the continuum limit extrapolation, as
we only have results at one lattice spacing. However, we know there is no
$O(a)$ error due to chiral symmetry~\cite{knn97} and the $O(a^2)$ and $O(m^2
a^2)$ errors are small~\cite{liu02,dll00,ddh02}. We shall use the results from
the small volume study in Ref.~\cite{dll00} to gain some feeling for the
magnitude of the $O(a^2)$ errors. We have calculated pion and rho masses on the
$6^3 \times 12, 8^3 \times 16$, and $10^3 \times 20$ lattices with the Wilson
gauge action at $\beta = 5.7, 5.85$, and 6.0 respectively. The figures of
$m_{\pi}a/\sqrt{\sigma}a$ and $m_{\rho}a/\sqrt{\sigma}a$ ($\sigma$ is the
string tension) vs. $\sigma a^2$ in Ref.~\cite{dll00} show that both of them are 
quite flat at $m_{\pi}/m_{\rho} = 0.4, 0.5$, and 0.6. Since the lattice spacing at 
$\beta = 5.7$, as measured with the Sommer $r_0$, is $0.171$ fm which is very close to
our present lattice whose lattice spacing as determined from $r_0$ is 0.175 fm,
we shall use the $\beta = 5.7$ data as the estimate for the present lattice.
After extrapolating the pion mass from $\beta = 5.7, 5.85$, and 6.0 in
Ref.~\cite{dll00} to the continuum limit, we find the central values of
$m_{\pi}$ at $\beta = 5.7$ are changed by 1 -- 3 \% for the cases with
$m_{\pi}/m_{\rho} = 0.4, 0.5$, and 0.6. We thus tentatively suggest that our
present pion masses are subject to the similar level of $O(a^2)$ errors due to the
continuum limit extrapolation. This, of course, needs to be checked with
calculations at smaller lattice spacings than what we have now.

\section{Chiral behavior of $f_{\pi}$ and $f_P$}  \label{f_pi}

\begin{figure}[t]
\includegraphics[width=0.8\hsize]{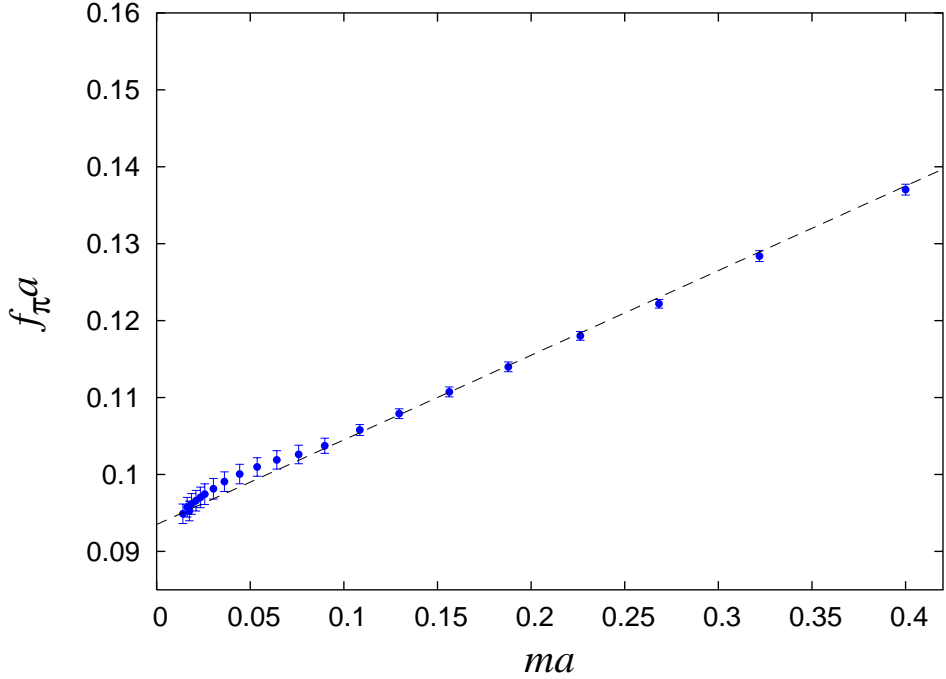}
\caption{\label{f_pi_long} $f_{\pi}$ determined from the the pseudoscalar
correlator $G_{PP}(\vec{p}= 0, t)$ in Eq.~(\ref{f_pi_eq}) as a function of the
quark mass $m$.  The dashed curve is a linear fit to the high mass region, $m =
0.1085$ ($m_{\pi} = 0.4438(36)$ or $438(7)\,{\rm MeV}$) to 0.4 ($m_{\pi} =
0.8740(30)$ or $862(13)\,{\rm MeV}$).}
\end{figure}

\begin{figure}[t]
\includegraphics[width=0.8\hsize]{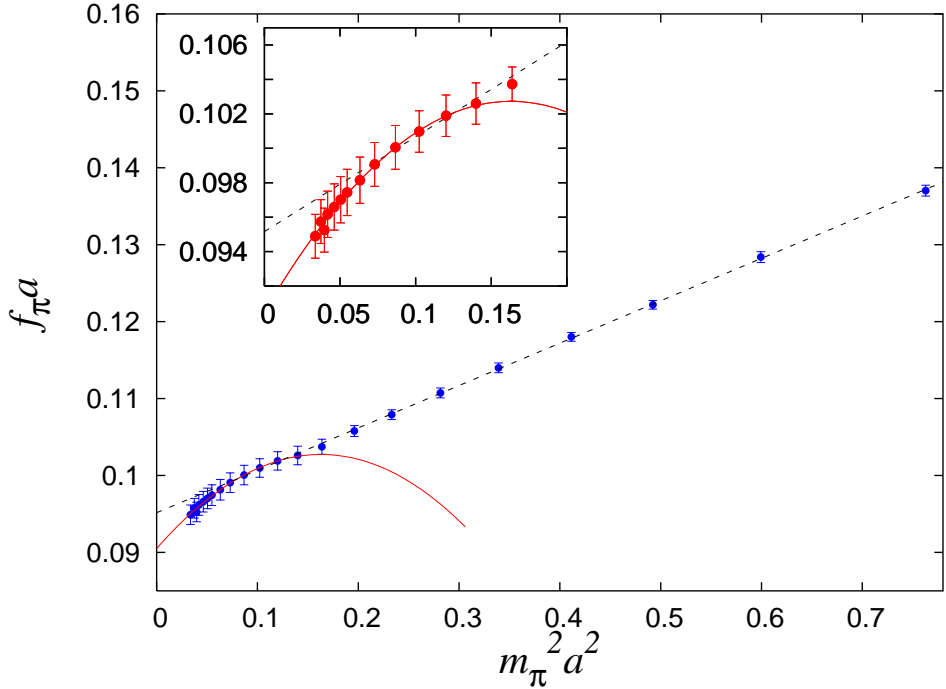}
\caption{\label{f_pi_mpi2} $f_{\pi}$ as a function of $m_{\pi}^2$. The dashed
curve is a fit with $f_{\pi}(0) + C_1 m_{\pi}^2$ to the high mass region,
$m_{\pi} = 0.4438(36)$ or $438(7)\,{\rm MeV}$ to 0.8740(30) or $862(13)\,{\rm
MeV}$).  The solid curve is the fit with Eq. (\ref{f_pi_mpi}) for the small
mass region, $m_{\pi} = 0.1844(71)$ or $182(8)\,{\rm MeV}$) to $m_{\pi, max} =
0.3748(38)$ or $370(7)\,{\rm MeV}$).  The insert is a blowup for the small mass
region.}
\end{figure}

\begin{figure}[ht]
\includegraphics[width=0.8\hsize]{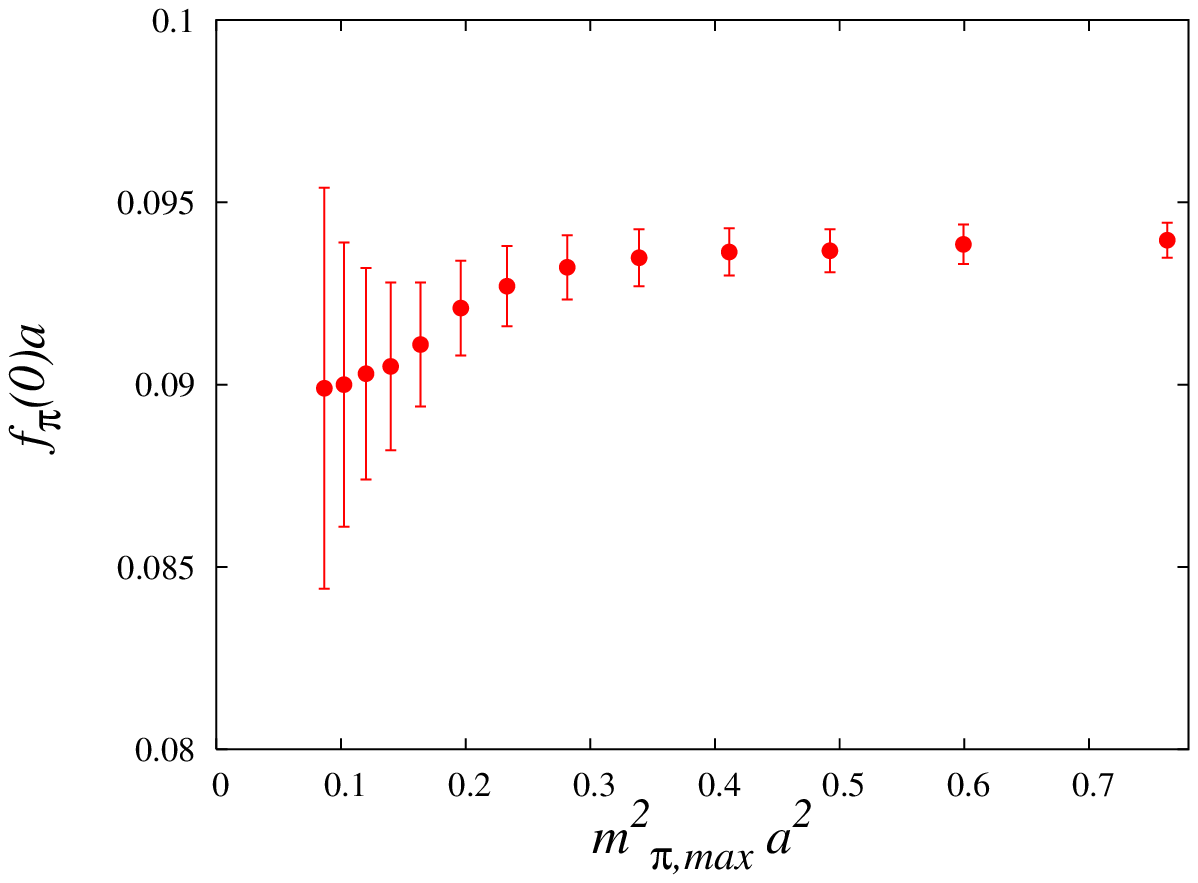}
\caption{\label{f_pi(0)}
$f_{\pi}(0)$ as determined from the fit Eq.~(\ref{f_pi_mpi}) as a function of 
the maximum pion mass $m_{\pi, max}$.}
\end{figure}

We show in Fig.~\ref{f_pi_long} our data on renormalized $f_{\pi}$ as
obtained~\cite{ddh02} from
\begin{equation} \label{f_pi_eq}
f_{\pi} = \lim_{t \longrightarrow \infty}\frac{2 m\, \sqrt{G_{PP}(\vec{p}= 0, t)
\,m_{\pi}} e^{m_{\pi}t/2}}{(m_{\pi})^2}.
\end{equation}

We see that if one fits a linear curve in quark mass from $0.1085$ ($m_{\pi} =
0.4438(36)$ or $438(7)\,{\rm MeV}$) to $0.4$ ($m_{\pi} = 0.8740(30)$ or
$862(13)\,{\rm MeV}$) as is usually done in the literature, the fit gives $
f_\pi = 0.0935(6) + 0.110(2)\,m$ with $\chi^2/NDF = 1.0$.  In this case, the
data below \mbox{$m$ = 0.1085} shows a hump with a one-sigma deviation from the
linear fit at the low mass end. Even though $f_{\pi}$ in full QCD with
dynamical fermions has a leading order chiral log~\cite{lp73,gl85}\, in the
form $m_{\pi}^2 \log (m_{\pi}^2)$ with a known coefficient, it is absent in the
quenched theory~\cite{sha92}. We interpret the bump as a consequence of the quenched
chiral log in the pion mass. Thus, we plot the the $f_{\pi}$ as a function of
$m_{\pi}^2$ in Fig. \ref{f_pi_mpi2} and fit with the following form
\begin{equation}  \label{f_pi_mpi}
f_{\pi} = f_{\pi}(0) + C_1 m_{\pi}^2 + C_2 m_{\pi}^4.
\end{equation}


\begin{figure}[t]
\includegraphics[width=0.8\hsize]{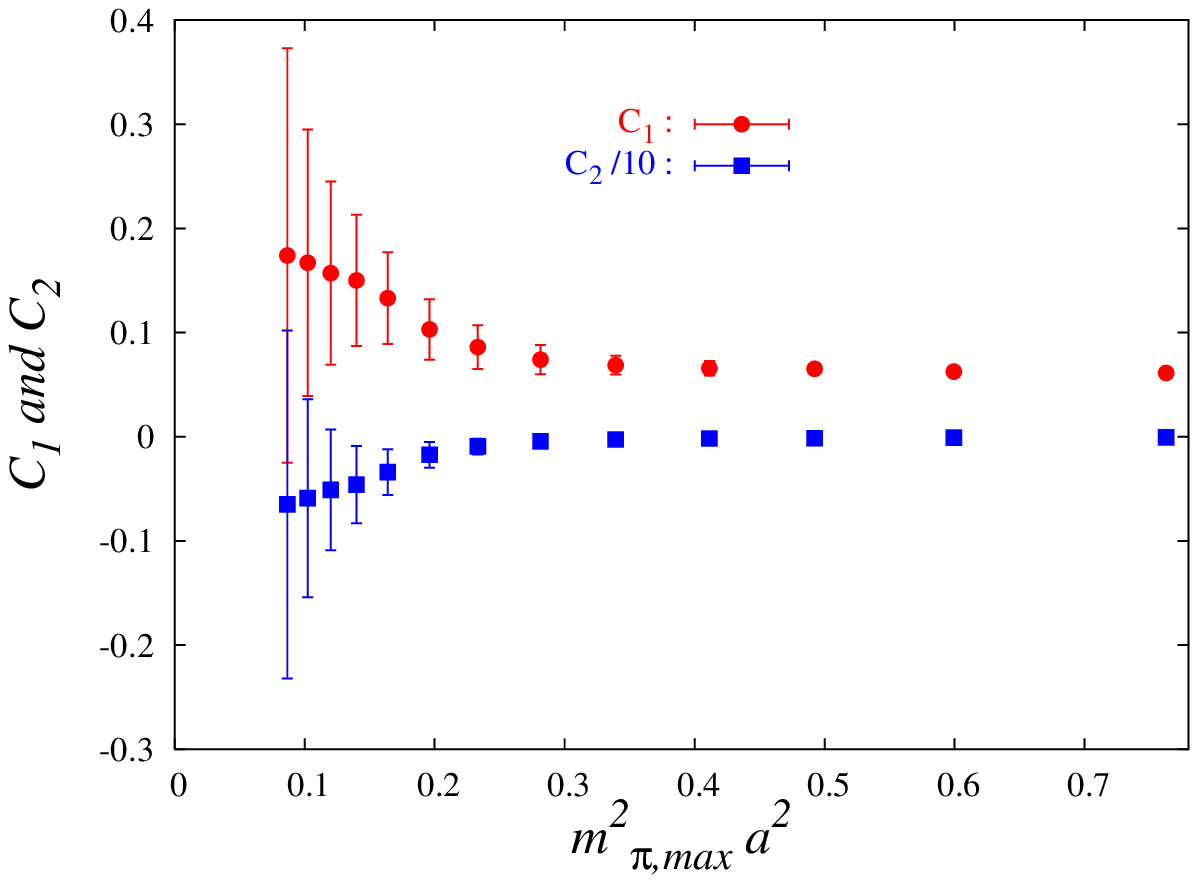}
\caption{\label{f_pi_C1_C2}
Same as in Fig.~\ref{f_pi(0)} but for $C_1$ and $C_{2}/10$.}
\end{figure}

%
%

As in the fit of $m_{\pi}^2$, we fix the minimum pion mass at 0.1844(71) and
fit the data to the above formula with variable ranges.  The results of
$f_{\pi}(0)$ are plotted in Fig.~\ref{f_pi(0)} as a function of the maximum
pion mass $m_{\pi, max}^2$.  We plot the same for $C_1$ and $C_{2}/10$ in
Fig.~\ref{f_pi_C1_C2}. Even though there appear to be some changes in the
region of $m_{\pi, max}$ between $0.3748(38)$ and $0.4840(32)$ for $f_{\pi}(0),
C_1$ and $C_2$, the errors are such that one cannot be as certain as in the
case of the chiral log $\delta$.  Thus, we shall quote the results for $m_{\pi,
max} = 0.3748(38)$ or $370(7)\,{\rm MeV}$.  They are given in
Table~\ref{f_pi_p}.  The fit for $f_{\pi}$ in the range up to $m_{\pi, max} =
0.3748(38)$ with the parameters in Table~\ref{f_pi_p} is plotted in
Fig.~\ref{f_pi_mpi2} as the solid lines for both the full figure and the
insert. We also fit $f_{\pi}$ with only the quadratic pion mass term for the
large pion mass region from $m_{\pi} = 0.4438(36)$ or $438(7)\,{\rm MeV}$ to
0.8740(30) or $862(13)\,{\rm MeV}$) and obtain $f_{\pi} = 0.0952(6) +
0.0551(13) m_{\pi}^2$. This fit is plotted as the dash line in
Fig. \ref{f_pi_mpi2}.

According to the quenched chiral perturbation theory~\cite{bde00}, $C_1 =
\frac{4 L_5}{f_{\pi}(0)}$. From our data reported in Table \ref{f_pi_p}, we
obtain $L_5 = 3.6(1.5) \times 10^{-3}$. This is comparable to $L_5 = 2.5(5) \times
10^{-3}$ as obtained from the study of $f_{\pi}$ with the clover fermion action
on Wilson gauge configurations at $\beta = 5.7$.  But it is substantially
larger than the phenomenological value of $1.4(5) \times 10^{-3}$~\cite{pic95}.

Using the experimental value $f_{\pi}(m_{\pi}) = 92.4\,{\rm MeV}$ and $m_{\pi}
= 137\,{\rm MeV}$ to set the scale, we determine the lattice spacing $a$ to be
$0.200(3)\,{\rm fm}$.  We see that $f_{\pi}(m_{\pi})$ is higher than
$f_{\pi}(0)$ by $\sim 5$\%.

\begin{table}[b]
\begin{center}
\caption{\label{f_pi_p} $f_{\pi}(0)$, $C_1$, and $C_{2}$ as fitted from
$m_{\pi} = 0.1844(71)$ or $182(8)\,{\rm MeV}$ to $m_{\pi} = 0.3748(38)$ or
$370(7)\,{\rm MeV}$.}
\vspace*{0.2in}
\begin{tabular}{c@{\hspace{2em}}c@{\hspace{2em}}c@{\hspace{2em}}c}
\hline
$f_{\pi}(0)$ &$C_1$&$C_{2}$ & $L_5$\\
 \hline
  0.0905(23) & 0.150(63)  & -0.46(37) & $3.6(1.5) \times 10^{-3}$ \\
   \hline
\end{tabular}
\end{center}
 \end{table}

The unrenormalized pseudoscalar decay constant of the pion is defined as
\begin{equation}  \label{f_P}
f_P^U = \langle 0|\bar{\psi} i \gamma_5 (1 - D/2) \psi |\pi(\vec{p} = 0)\rangle 
\end{equation}
From the chiral Ward identity, we have
\begin{equation}  \label{awi}
Z_A\partial_{\mu} A_{\mu} = 2 Z_m m Z_P P,
\end{equation}
where $A_{\mu}=\bar{\psi}i\gamma_{\mu}\gamma_5(1 - D/2)\psi$ and $P =\bar{\psi}
i\gamma_5(1 - D/2)\psi$.  Since $Z_m = Z_S^{-1}$ and $Z_S = Z_P$ due to the
fact the scalar density $S=\bar{\psi}(1 - D/2)\psi$ and the pseudoscalar
density $P$ are in the same chiral multiplet, $Z_m$ and $Z_P$ cancel in
Eq.~(\ref{awi}).  Therefore, from the relation of the on-shell matrix elements
\begin{equation}
\langle 0|Z_A\partial_4 A_4|\pi(\vec{p} = 0)\rangle = 2 m \langle 0|P|\pi(\vec{p} = 0)\rangle,
\end{equation}
one obtains that 
\begin{equation}  \label{f_P_unren}
\frac{f_P^U}{f_{\pi}} = \frac{m_{\pi}^2}{2m},
\end{equation}
where $f_{\pi}$ is renormalized while $f_P^U$ and $m$ are unrenormalized.  The
relation also holds when both $f_P$ and $m$ are renormalized since $Z_m =
Z_P^{-1}$.  It is clear from Eq.~(\ref{f_P_unren}) that the chiral behavior of
$f_P^U$ is the combination of those of $m_{\pi}^2/2m$ and $f_{\pi}$.

\section{Quenched $m_{N}$}

\begin{figure}[ht]
\includegraphics[width=0.8\hsize]{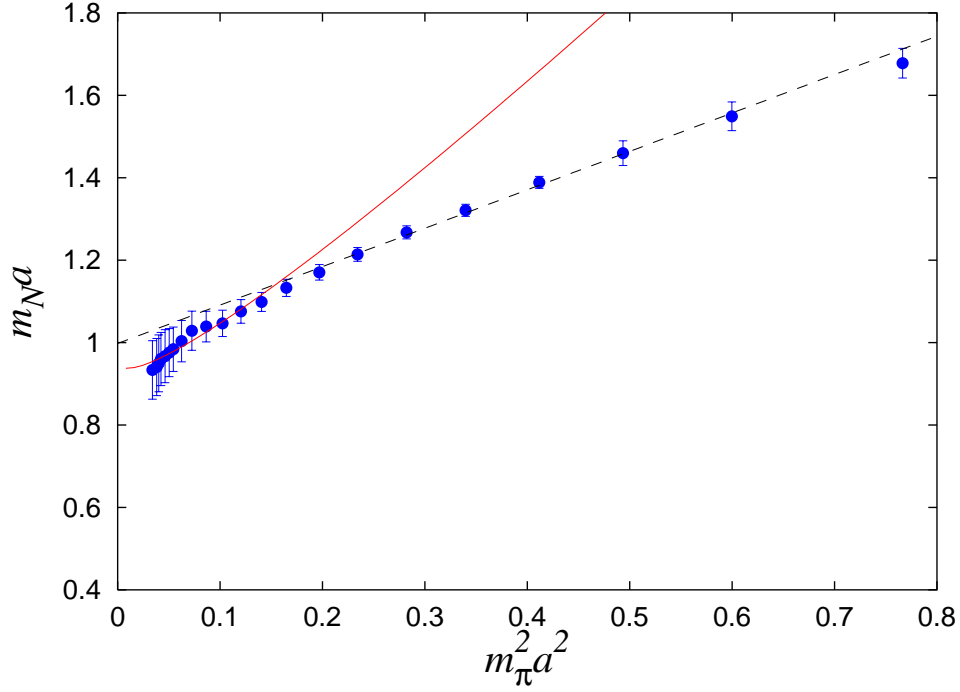}
\caption{\label{m_N_long} The nucleon mass as a function of $m_{\pi}^2$.  The
dashed line is a fit only linear in $m_{\pi}^2$ for the range of $m_{\pi}$ from
$438(7)\,{\rm MeV}$ to $862(13)\,{\rm MeV}$.  The solid line is a fit with
Eq.~(\ref{N_fit}) for the range $m_{\pi} = 182(8)\, {\rm MeV}$ to $316(6)\,{\rm
MeV}$.}
\end{figure}

We first plot the nucleon mass as a function of $m_{\pi}^2$ in
Fig.~\ref{m_N_long}.  We see that it is fairly linear in $m_{\pi}^2$ all the
way down to $m_{\pi} \sim 0.4438(36)$ ($438(7)\,{\rm MeV}$).  Below that point,
there appears to be a deviation from the linear behavior.  Following the case
of $f_{\pi}$, we shall first fit the $N$ mass only linear in $m_{\pi}^2$ in the
range of $m_{\pi}^2 = 0.196 - 0.763$ and obtain $m_N = 0.998(21) + 0.932(74)
m_{\pi}^2$.  This fit is represented by a dashed line in Fig.~\ref{m_N_long}.
We see that the data (in the small $m_{\pi}$ region) in this case are
systematically lower than the linear fit.

\begin{figure}[ht]
\includegraphics[width=0.8\hsize]{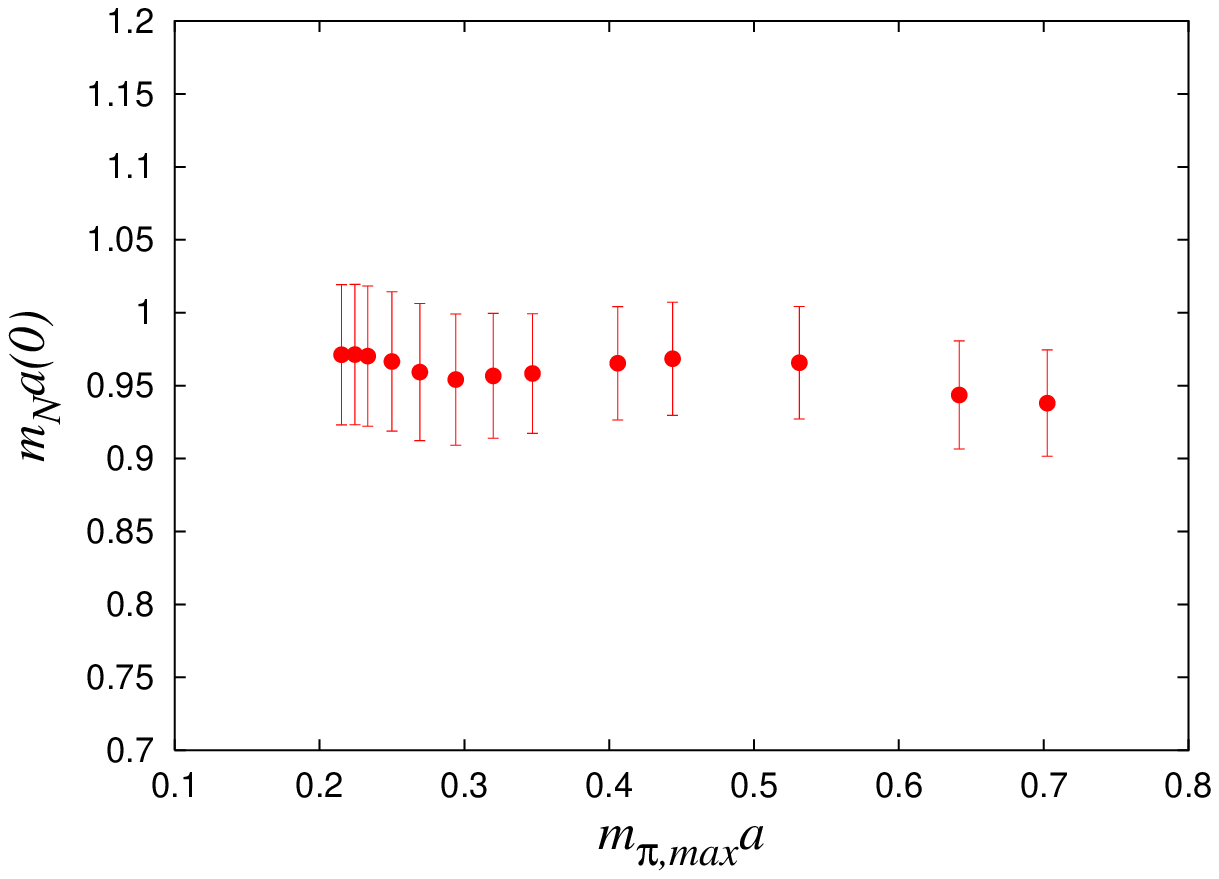}
\caption{\label{m_N_0}
Fitted $m_N(0)$ in Eq.~(\ref{N_fit}) as a function of $m_{\pi,max}$ of the
fitting range.}
\end{figure}

According to the quenched $\chi$PT for the baryons~\cite{ls96}, the one-loop
result predicts the presence of the non-analytic terms $m_{\pi}$ and $m_{\pi}^2
\log (m_{\pi})$, both of which are proportional to the quenched $\delta$.
Thus, we fit the nucleon data from the lowest pion mass, i.e.\ $m_{\pi, min} =
182(8)\,{\rm MeV}$ upward as done in previous sections with the form
\begin{equation}  \label{N_fit}
m_N = m_N(0) + C_{1/2} m_{\pi} + C_1 m_{\pi}^2 + C_{1L} m_{\pi}^2
\log (\frac{m_{\pi}}{\Lambda_\chi}) + C_{3/2} m_{\pi}^3 ,
\end{equation}
We should mention again that the fit does not depend on the value of
$\Lambda_{\chi}$, since there is an analytic term in $m_{\pi}^2$.  We take
$\Lambda_{\chi}=1.16$ for this case.

\begin{table}[ht]
\begin{center}
\caption{\label{p_nucleon} $m_N(0)$, $C_{1/2}$, $C_1$, $C_{1L}$, and $C_{3/2}$
as fitted from $m_{\pi} = 0.1844(71)$ ($182(8)\,{\rm MeV}$) to $0.3200(43)$
($316(6)\,{\rm MeV}$).}
\vspace*{0.2in}
\begin{tabular}{c@{\hspace{2em}}c@{\hspace{2em}}c@{\hspace{2em}}c@{\hspace{2em}}c}
\hline
$m_N(0)$&$C_{1/2}$&$C_1$&$C_{1L}$&$C_{3/2}$\\
\hline
  0.957(43) & -0.358(234) & 2.44(67)  & 0.319(290) & 0.016(199) \\
\hline
\end{tabular}
\end{center}
\end{table}

\begin{figure}[ht]
\includegraphics[width=0.8\hsize]{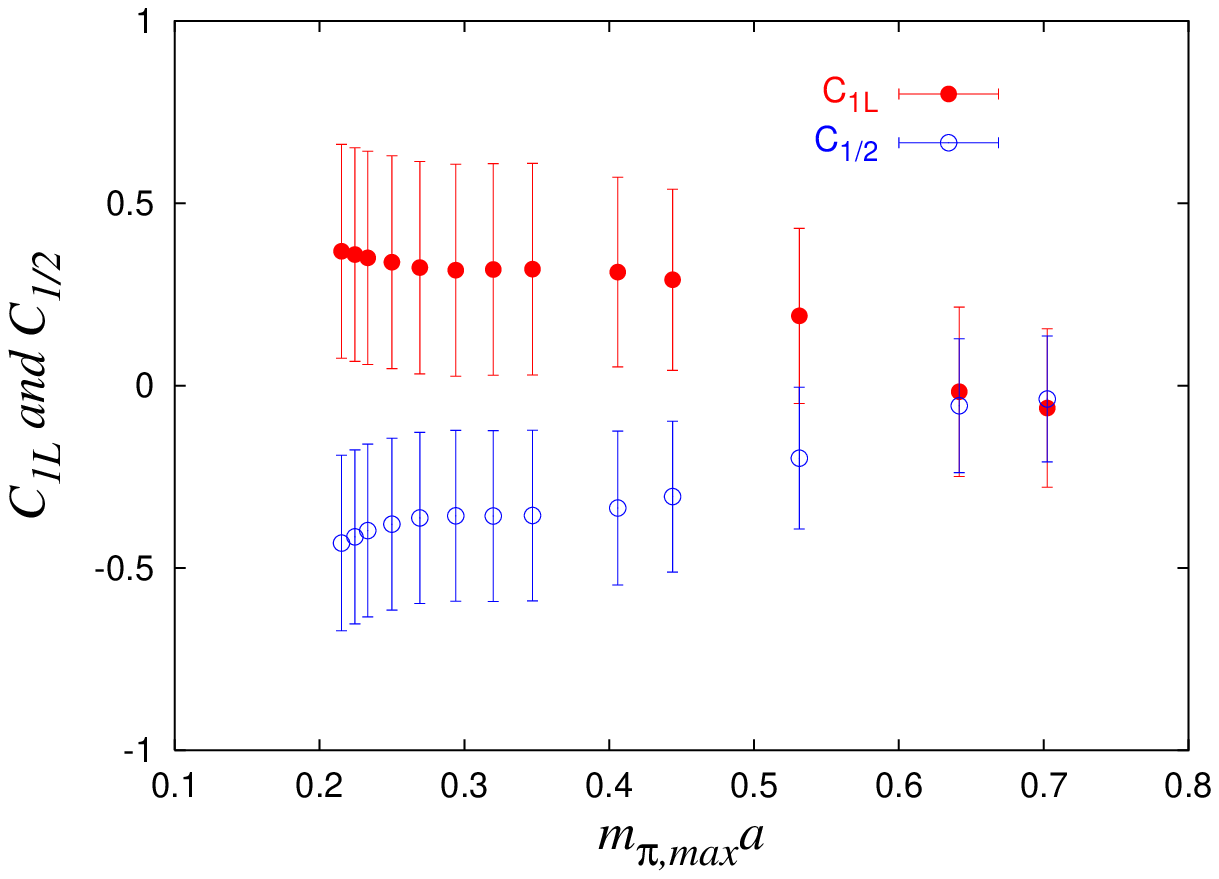}\\
\includegraphics[width=0.8\hsize]{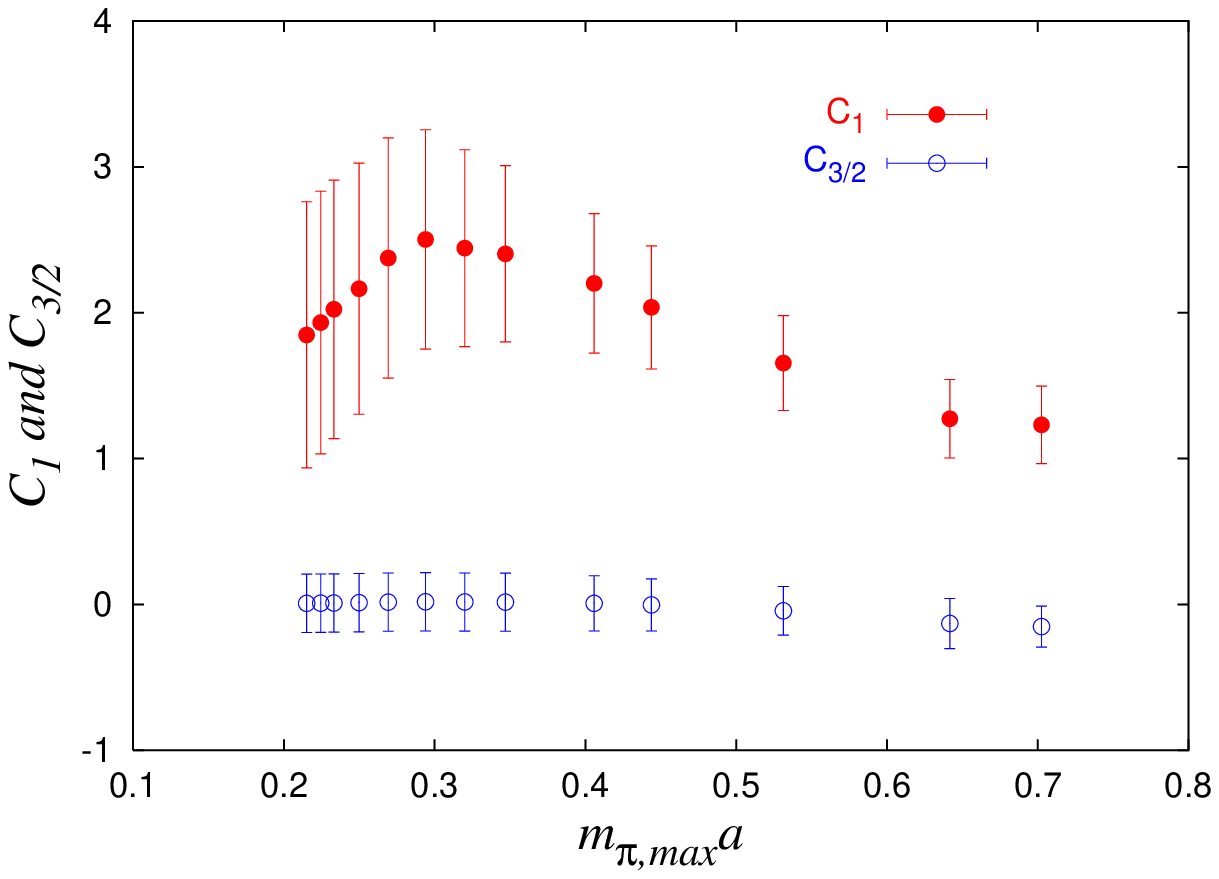}
\caption{\label{nucleon_C1L} The same as in Fig.~\ref{m_N_0} for the
coefficients $C_{1L}$ and $C_{1/2}$ (upper figure) and for the coefficients
$C_1$ and $C_{3/2}$ (lower figure).}
\end{figure}

\begin{figure}[ht]
\includegraphics[width=0.8\hsize]{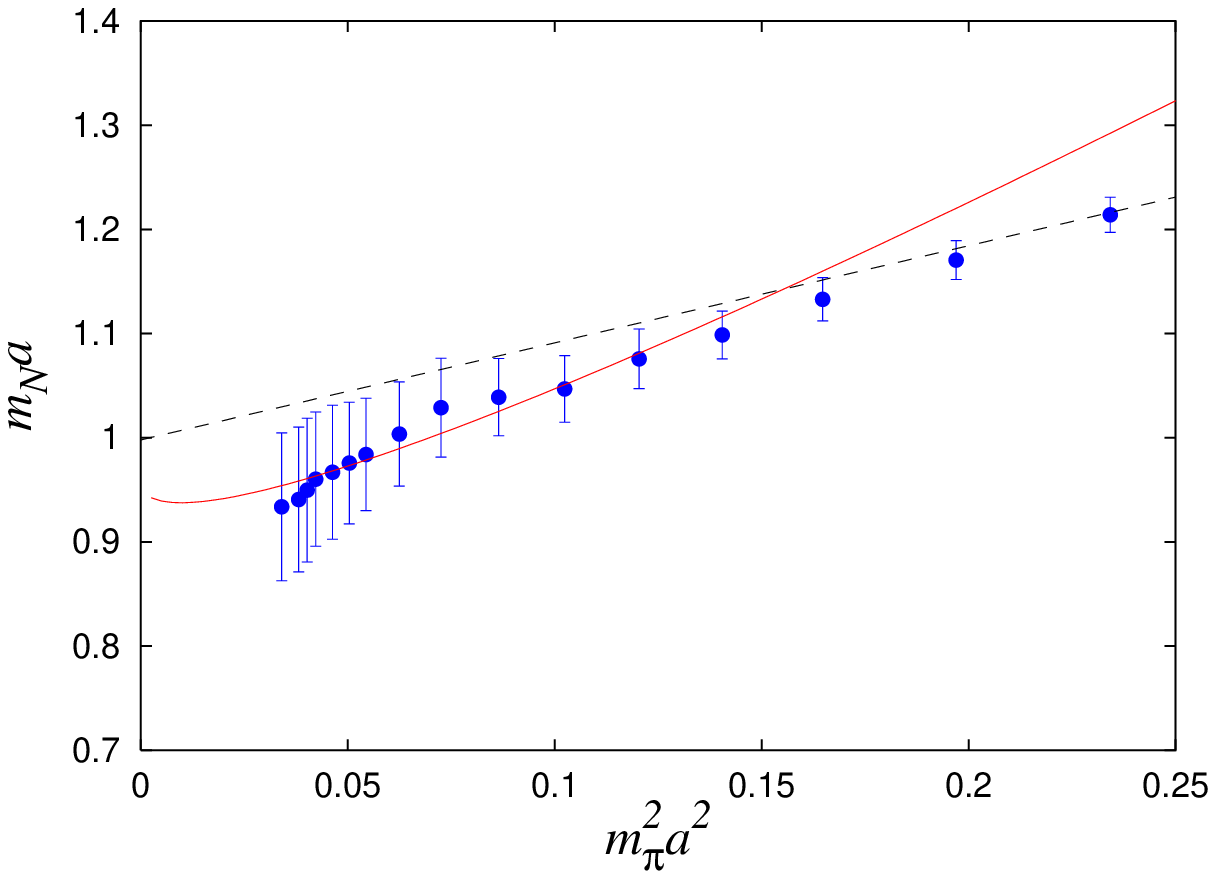}
\caption{\label{m_N_short} The nucleon mass as a function of $m_{\pi}^2$ for
smaller pion masses.  The solid line is a fit with Eq.~(\ref{N_fit}) for the
range $m_{\pi} = 0.1844(71)$ ($182(8)\,{\rm MeV}$) to $0.3200(43)$
($316(6)\,{\rm MeV}$).  The dashed line is a linear fit from a higher mass
range of $m_{\pi} = 0.4438(36)$ ($438(7)\,{\rm MeV}$) to $0.8740(30)$
($862(13)\,{\rm MeV}$).}
\end{figure}

The fitted results of $m_N(0)$ are plotted in Fig.~\ref{m_N_0} as a function of
the maximum pion mass $m_{\pi, max}$ of the fitted range.  We see that it is
rather stable in the range fitted.  Similarly, the coefficients $C_{1/2}$,
$C_{1L}$, $C_1$ and $C_{3/2}$ are plotted in Fig.~\ref{nucleon_C1L}.  We see
that both $C_{1/2}$ and $C_{1L}$ have rather smooth behavior in $m_{\pi, max}$
from $0.3$ to $0.55$ and approach zero around $m_{\pi, max} = 0.6$ -- $0.7$.
On the other hand, $C_1$ has more of a variation in the range $m_{\pi, max} =
0.3 -0.55$.  When $m_{\pi, max}$ reaches the region $0.6$ -- $0.7$, the error
bars do not touch those at $m_{\pi, max} \sim 0.3$.  We note that $C_{3/2}$ is
consistent with zero in the whole range of the fit.  We have tried to include
the next analytic term $m_{\pi}^4$~\cite{leinweber03}.  It turns out that
$m_N(0), C_{1/2}$, and $C_1$ are not changed outside their errors in this case.
However, $C_{3/2}$ and $C_2$ with the $m_{\pi}^4$ term become ill-determined.
They are consistent with zero but with large errors.  Thus, the data seem to
suggest that there is no evidence for a large $m_{\pi}^3$ term in the quenched
nucleon mass.  We quote the fit evaluated at $m_{\pi, max} = 0.3200(43)$
($316(6)\,{\rm MeV}$) in Table~\ref{p_nucleon}.  This fit is shown as the solid
curve in Fig.~\ref{m_N_short} for $m_{\pi}^2$ up to $0.4840(32)$ ($m_{\pi} =
478(8)\,{\rm MeV}$).  Also shown in the figure as the dashed curve is the
linear fit from the higher pion mass range of $m_{\pi} = 0.4438(36)$
($438(7)\,{\rm MeV}$) to $0.8740(30)$ ($862(13)\,{\rm MeV}$).

We should mention that even though the error is large, $C_{1/2}$ is consistent
with that predicted by the one-loop $\chi$PT~\cite{ls96}, for which $C_{1/2} =
- \frac{3 \pi}{2} (D- 3F)^2 \delta$.  Using $\delta = 0.24(3)$ from our fit in
Sec.~\ref{chi_log_mpi}, it gives $C_{1/2} = -0.37(5)$.  This is consistent with
the value in Table~\ref{p_nucleon} which is $-0.358(234)$.  $C_1 = 2.44(67)$
from the low mass fit is much larger than that of the high mass fit where $C_1
= 0.932(74)$.

\section{Conclusion}

The chiral extrapolation has always been a major challenge in lattice
QCD calculations, primarily due to the fact that one does not know if
the formula used to fit is correct in the range of the available data.
The idea of using $\chi$PT to guide the fit is useful and appropriate
except it is complicated by the fact that one does not know the range
in where the one-loop formula is valid.  In this work, we attempt to
answer this question by fitting the data from our smallest pion mass
(close to the physical value) upward and seeing where the low-energy
parameters change.  This is commensurate with the $\chi$PT approach
which expands around the small pion mass and momentum.  The chiral log
$\delta$ from fitting the pion mass to the quenched one-loop chiral
formula clearly demonstrates a sharp transition around $300-400\,{\rm
MeV}$ in the pion mass.  We take this as an evidence to reveal the
upper limit for the applicability of the one-loop formula in quenched
$\chi$PT\@. We have also fitted the pion mass with the cactus
re-summed diagrams in quenched $\chi$PT and the results suggest that
its range of applicability extends beyond the one-loop formula. We
should point out that fitting with the cactus re-summed diagrams is
preferred over the one-loop formula since it gives a scale-independent
$\delta$. However, not all the physical quantities have cactus
re-summed results in the $\chi$PT and in many cases, one would rely on
one-loop formula to do the chiral fitting. Thus, it is essential to
find out where the one-loop formula is valid.  To be sure that this is
not skewed by the finite volume effects, we compared our data on the
$16^3 \times 28$ lattice against those on the $12^3 \times 28$ lattice
and found that the finite volume errors are much smaller than the
statistical errors.  Although less certain due to larger errors, the
analytic term in the $N$ mass and the non-leading chiral log in
$f_{\pi}$ show variations in the pion mass range of $300-400\,{\rm
MeV}$, and the $C_1$ coefficients change by a factor $\sim 2 - 3$
compared with those from fits involving exclusively pion masses beyond
this range.  Thus we tentatively conclude that the one-loop quenched
$\chi$PT is valid for pion mass less than $\sim 300\,{\rm MeV}$.  It
is claimed that using a different regularization scheme such as the
finite range regularization~\cite{lty03} will extend the range of
validity for the one-loop quenched $\chi$PT from that of the
dimensional regularization, but it remains to be checked by lattice
Monte Carlo data at low pion mass.

One of the valuable lessons learned in this study is that it seems futile to
extrapolate to the chiral region from pion mass beyond $\sim 300\,{\rm MeV}$.
We tried to fit the pion mass with the one-loop formula by extrapolating from
$m_{\pi} = 862(13)\,{\rm MeV}$ and found that, no matter how small pion masses
one includes, $\delta$ is inevitably different (by more than two sigmas) from
that obtained with the data limited to $m_{\pi} < \sim 250\,{\rm MeV}$.  One
could try to model the functional form at higher masses, but it is not likely
to succeed, because one does not know {\it a priori\/} how and when the modeled
formula eases into that of the one-loop chiral $\chi$PT\@.  Direct simulation
at pion masses less than $\sim 300\,{\rm MeV}$ is nearly impossible (or at the
very least an extremely difficult task) for lattice calculations employing
fermion actions which suffer from exceptional configurations and/or critical
slowing down.  An overlap-fermion calculation is numerically intensive, but
with approximately fixed cost -- its critical slowing down is rather
mild~\cite{dll00}.  The fact that it has chiral symmetry alleviates the worry
that something may go wrong at small quark masses.  Furthermore, the chiral
behavior study is greatly enhanced by the incorporation of the multi-mass
algorithm which facilitates the constrained curve-fitting method to discern the
log structure.  One does not know how different the chiral behavior in full QCD
will be from that of the quenched approximation, but it is hard to imagine that
there are different scales for the sea and valence quarks.  If so, one will
need to work harder to lower the dynamical quark masses such that the
corresponding pion mass is below $\sim 300\,{\rm MeV}$.

\section{Acknowledgment}

This work is partially supported by U.S. DOE Grants DE-FG05-84ER40154 and
DE-FG02-95ER40907.  We thank A. Alexandru, C. Bernard, M. Chanowitz, N. Christ,
M. Golterman, A. Kennedy, D. Leinweber, S. Sharpe, A. Soni, S. Tamhankar,
H. Thacker, and H. Wittig for useful discussions and suggestions.  Thanks are
also due to Y. Aoki, K. Holland, and C. Gattringer for providing us unpublished
numbers.


\begin{thebibliography}{99}
\bibitem{note1} 
We will postpone the consideration of the finite volume effect and zero mode
contribution to Section~\ref{fv}.
\end{thebibliography}
\end{document}